\PassOptionsToPackage{unicode}{hyperref}
\PassOptionsToPackage{hyphens}{url}
\PassOptionsToPackage{dvipsnames,svgnames,x11names}{xcolor}

\documentclass[
  12pt]{article}

\usepackage{amsmath,amssymb}
\usepackage{iftex}
\ifPDFTeX
  \usepackage[T1]{fontenc}
  \usepackage[utf8]{inputenc}
  \usepackage{textcomp} 
\else 
  \usepackage{unicode-math}
  \defaultfontfeatures{Scale=MatchLowercase}
  \defaultfontfeatures[\rmfamily]{Ligatures=TeX,Scale=1}
\fi
\usepackage{lmodern}

\ifPDFTeX\else  
    
\fi

\IfFileExists{upquote.sty}{\usepackage{upquote}}{}
\IfFileExists{microtype.sty}{
  \usepackage[]{microtype}
  \UseMicrotypeSet[protrusion]{basicmath} 
}{}
\makeatletter
\@ifundefined{KOMAClassName}{
  \IfFileExists{parskip.sty}{
    \usepackage{parskip}
  }{
    \setlength{\parindent}{0pt}
    \setlength{\parskip}{6pt plus 2pt minus 1pt}}
}{
  \KOMAoptions{parskip=half}}
\makeatother
\usepackage{xcolor}
\makeatletter
\ifx\paragraph\undefined\else
  \let\oldparagraph\paragraph
  \renewcommand{\paragraph}{
    \@ifstar
      \xxxParagraphStar
      \xxxParagraphNoStar
  }
  \newcommand{\xxxParagraphStar}[1]{\oldparagraph*{#1}\mbox{}}
  \newcommand{\xxxParagraphNoStar}[1]{\oldparagraph{#1}\mbox{}}
\fi
\ifx\subparagraph\undefined\else
  \let\oldsubparagraph\subparagraph
  \renewcommand{\subparagraph}{
    \@ifstar
      \xxxSubParagraphStar
      \xxxSubParagraphNoStar
  }
  \newcommand{\xxxSubParagraphStar}[1]{\oldsubparagraph*{#1}\mbox{}}
  \newcommand{\xxxSubParagraphNoStar}[1]{\oldsubparagraph{#1}\mbox{}}
\fi
\makeatother

\usepackage{longtable,booktabs,array}
\usepackage{calc} 
\usepackage{etoolbox}
\usepackage{multirow}
\makeatletter
\patchcmd\longtable{\par}{\if@noskipsec\mbox{}\fi\par}{}{}
\makeatother

\IfFileExists{footnotehyper.sty}{\usepackage{footnotehyper}}{\usepackage{footnote}}
\makesavenoteenv{longtable}
\usepackage{graphicx}
\makeatletter
\def\maxwidth{\ifdim\Gin@nat@width>\linewidth\linewidth\else\Gin@nat@width\fi}
\def\maxheight{\ifdim\Gin@nat@height>\textheight\textheight\else\Gin@nat@height\fi}
\makeatother

\setkeys{Gin}{width=\maxwidth,height=\maxheight,keepaspectratio}

\makeatletter
\def\fps@figure{htbp}
\makeatother

\makeatletter
\@ifpackageloaded{caption}{}{\usepackage{caption}}
\AtBeginDocument{
\ifdefined\contentsname
  \renewcommand*\contentsname{Table of contents}
\else
  \newcommand\contentsname{Table of contents}
\fi
\ifdefined\listfigurename
  \renewcommand*\listfigurename{List of Figures}
\else
  \newcommand\listfigurename{List of Figures}
\fi
\ifdefined\listtablename
  \renewcommand*\listtablename{List of Tables}
\else
  \newcommand\listtablename{List of Tables}
\fi
\ifdefined\figurename
  \renewcommand*\figurename{Figure}
\else
  \newcommand\figurename{Figure}
\fi
\ifdefined\tablename
  \renewcommand*\tablename{Table}
\else
  \newcommand\tablename{Table}
\fi
}
\@ifpackageloaded{float}{}{\usepackage{float}}
\floatstyle{ruled}
\@ifundefined{c@chapter}{\newfloat{codelisting}{h}{lop}}{\newfloat{codelisting}{h}{lop}[chapter]}
\floatname{codelisting}{Listing}

\makeatother
\makeatletter
\@ifpackageloaded{caption}{}{\usepackage{caption}}
\@ifpackageloaded{subcaption}{}{\usepackage{subcaption}}
\makeatother

\ifLuaTeX
  \usepackage{selnolig}  
\fi
\usepackage[]{natbib}
\usepackage{bookmark}

\IfFileExists{xurl.sty}{\usepackage{xurl}}{} 
\hypersetup{
  pdftitle={Title},
  pdfauthor={Author 1; Author 2},
  pdfkeywords={3 to 6 keywords, that do not appear in the title},
  colorlinks=true,
  linkcolor={blue},
  filecolor={Maroon},
  citecolor={Blue},
  urlcolor={Blue},
  pdfcreator={LaTeX via pandoc}}

\usepackage{bm}

\newcommand{\bs}{\textbf{s}}

\newcommand{\anon}{1}

\begin{document}

\def\spacingset#1{\renewcommand{\baselinestretch}
{#1}\small\normalsize} \spacingset{1}

\if1\anon
{
  \title{\bf Double zero-inflated spatio-temporal modeling of daily precipitation under detection thresholds}
  
  \author{Juan Marcen-Gutierrez\textsuperscript{1,}\thanks{Corresponding author: \texttt{jmarcen@unizar.es}}, 
          Jorge Castillo-Mateo\textsuperscript{1}, 
          Alan E. Gelfand\textsuperscript{2}, \\
          Jesús Asín\textsuperscript{1}, and 
          Ana C. Cebrián\textsuperscript{1} \\[2ex]
          \small\textsuperscript{1}Department of Statistical Methods and IUMA, University of Zaragoza \\
          \small\textsuperscript{2}Department of Statistical Science, Duke University}
  \date{}
  \maketitle
} \fi

\if0\anon
{
  \bigskip
  \bigskip
  \bigskip
  \begin{center}
    {\LARGE\bf Title}
\end{center}
  \medskip
} \fi

\bigskip
\begin{abstract}
Explaining precipitation behavior at daily scale is important for fine scale understanding of the mechanisms driving precipitation.  However, this effort is challenging because of the frequent incidence of zeros.  The challenge is amplified by the acknowledged incidence of two types of zeros---absence of precipitation as a dry event and absence of measured precipitation due to detection limits.  In this work, we propose a multilevel spatio-temporal model which allows us to distinguish and explain the two types of zeros, as well as to model positive precipitation above the detection limit.  The methodology combines a point mass at zero with probability modeled through a probit regression, a Gamma regression for latent positive precipitation amounts, and an observation mechanism subject to threshold-induced censoring. To capture spatial dependencies, Gaussian processes are employed in each regression model.  Working within a Bayesian framework, we can obtain a rich range of inference with exact uncertainty.  In particular, we provide model-based inference tools to compare and quantify differences between the true precipitation process and its observed counterpart across relevant characteristics.  We apply our model to the analysis of daily spring observations at 70 sites over 15 years from the Ebro River Basin in northeastern Spain. Our findings indicate that the threshold strongly affects the occurrence of observed precipitation, especially in humid regions. While its impact on total accumulated amounts is small, it can exert a relevant effect on upper quantiles.
\end{abstract}

\vfill

\newpage
\spacingset{1.4} 

\section{Introduction}\label{Introduction}
Precipitation is an important driver of hydrological and environmental processes, yet its statistical modeling remains challenging due to its intermittent nature and strong skewness \citep{pendergrass2018}, as well as its complex dependence across space and time \citep{paschalis2013}. The impact of these characteristics is highly dependent on the temporal scale; while zero-precipitation events are often negligible in monthly \citep{marques2020} or annual \citep{sebille2017} series, they are a structural feature of daily data \citep{stauffer2017}. At this scale, recorded zeros do not always imply an absence of precipitation; devices typically have a detection threshold $\epsilon$, below which precipitation is unrecorded \citep{camuffo2022}. This left-censoring inflates the observed frequency of dry days and obscures the underlying mechanisms governing precipitation and wet-dry transitions \citep{stern1984}. 

This results in an observational model where an observed zero, in fact, incorporates two types of zeros.  One is a dry event, i.e., at the time and location of the measurement, the associated environmental conditions did not produce any precipitation.  The other is a censoring event; the amount of precipitation at that time and location was not zero but below the operating detection level of the measuring device. A natural way to address this issue is to define an underlying latent ``true'' precipitation process, in which precipitation arises from two distinct mechanisms: a binary process governing the occurrence of precipitation, and a continuous process governing its positive amount. In the absence of zeros, the positive component can be modeled using standard distributions for positive continuous data such as the Gamma distribution. When zeros are present, however, additional structure is required.  

The literature typically distinguishes between structural zeros (dry events) and positive outcomes (wet events) using different approaches. However, the distinction between dry event $0$'s and those induced by $\epsilon$ is not directly accounted for.
\cite{berrocal2008} propose a mixture-based spatial model, where occurrence is driven by a latent Gaussian process and the precipitation accumulation is assumed to follow a Gamma distribution.  
\cite{fernandes2009} introduce a methodology for zero-inflated (ZI) spatio-temporal models, which, when applied to precipitation data, the Gamma distribution yielded the best results. 
\cite{sun2015} model the precipitation occurrence using a threshold space-time $t$ random field. 
\cite{dzupire2018} model the occurrence using Poisson processes, while the intensity is modeled using Gamma distribution as the magnitude of the jumps of the process. 
\cite{johnson2023} combine radar data and observational data to forecast the precipitation field using a tobit model. 
\cite{euan2024} propose a hierarchical regime-based spatio-temporal model for precipitation data that identifies distinct regimes using neighboring site information, allowing spatial and temporal dependencies to vary between regimes through a Bayesian approach. 
\cite{zhang2024} propose a downscaling method combining the use of latent Gaussian models and stochastic partial differential equations in order to obtain high-resolution precipitation maps. 
While beyond the scope of this work, it is worth noting a recent surge in spatio-temporal extreme precipitation literature that also incorporates ZI modeling frameworks \citep[see, e.g.,][]{richards2023, vandeskog2025, zhong2025}. 
 
Measurement detection thresholds introduce an additional mechanism for generating observed zeros that remains largely overlooked in the literature. We note one example, \cite{berrocal2008} which recorded and modeled all precipitation accumulations below $0.01$ inches as zeros. Again, despite its ubiquity in precipitation data, this instrumental limitation remains largely unacknowledged and its impact is seldom discussed in most studies \citep{camuffo2022}.

In the ecological literature, we find work noting that observed zeros may arise from true absence or from \emph{imperfect detection} of a positive underlying process \citep{royle2003}.  In fact, some literature there proposes models that account for two types of zeros.  Specifically, in the context of percent biomass of tree species on plots, \cite{tang2023} and \cite{tang2025} propose a ZI beta distribution spatial model and a ZI tobit regression model, respectively. Both models explicitly account for two distinct sources of zeros: structural zeros (due to unsuitability), and zeros by chance (sampling variability), showing that incorporating both sources improves model performance.  However, the two types of zeros suffer from an identifiability challenge.  While they can be introduced into the model specification, there is nothing in the observed data to distinguish them.  Strong prior assumptions are needed to address this identifiability issue.

By contrast, with precipitation data, this mechanism is driven by specified measurement thresholds rather than stochastic detection.  In different words, an important implication of introducing a detection threshold is that the zero observations arise from two distinct and identifiable sources: true absence of precipitation and positive but undetected precipitation below a specified $\epsilon$. This distinction allows modeling to separate structural dry events from measurement-induced zeros, a feature that is not available in standard ZI models without explicit censoring. This serves as one of the primary contributions of this manuscript. Additionally, in many practical settings the detection limit is not constant, but varies across observed locations and time due to differences in instrumentation or measurement protocols. To accommodate this, we allow the threshold to vary across year, day within year, and location, yielding a flexible observation model that accounts for spatial and temporal heterogeneity in detection.

We note that there are other settings where our double zero-inflation with detection thresholds may be applicable.  One example is with environmental contaminants.  Atmospheric deposition (wet vs. dry) of sulfates and nitrates are of interest \citep{sahu2010}.  Minimum detection levels associated with these  pollutants are available for the associated measurement strategy. 
Another example concerns plant abundance at sites where abundance is measured with regard to biomass. There is a minimum detection level literature for biomass \citep{buxbaum2022, munser2025}.

More precisely, our contribution is to propose a hierarchical space-time model that explicitly incorporates these known detection limits, representing precipitation through two latent processes: a binary occurrence process modeled via a probit regression and a positive amount process modeled via a Gamma regression. The \emph{observed} data arise from a censored version of this latent \emph{true} process, whereby precipitation amounts below $\epsilon$ are recorded as zero. Both components are specified as spatially and temporally dependent processes, allowing for dependence across locations and over time. Model fitting and inference are conducted within a fully Bayesian hierarchical framework, enabling joint modeling of the latent occurrence, intensity, and observation processes. To implement this, we develop a custom Markov chain Monte Carlo (MCMC) algorithm that incorporates a double data augmentation step.

Our modeling approach supports both explanation and prediction, while propagating uncertainty from all sources, including latent processes, model parameters, and censoring. As a result, the proposed framework allows one to quantify the discrepancy between the true and observed precipitation processes in terms of incidence of both missed precipitation occurrences and undetected precipitation amounts, along with their associated uncertainty across space and time.  It further enables prediction, using various choices of $\epsilon$, of incidence of both types of zeros as well as positive precipitation amounts at arbitrary years, days, and locations within our study region.  Here, such prediction is facilitated by the introduction of spatial and temporal dependence.  Within this framework, we also implement necessary model adequacy and model comparison.

The value of a model that incorporates detection limits lies in its ability to estimate true precipitation amounts, including light events below the detection threshold. This enables the assessment of biases in precipitation amounts, specifically in the number of wet days. This, in turn, affects the number and length of wet spells and the estimation of drought indices, such as the Standardized Precipitation Index. The detection limit can also yield relevant biases in the estimation of precipitation quantiles. Furthermore, accurately estimating light precipitation or drizzle is important for sensitive ecosystems, as well as for evapotranspiration and infiltration models. Finally, the proposed model allows for an appropriate combination of stations and time periods with different detection limits, avoiding biases in analyses over time and across stations.

We analyze data consisting of observations from 70 daily precipitation series measured by tipping-bucket rain gauges located across the Ebro River Basin in northeastern Spain, covering from March to May from 2010 to 2024. The basin spans about 85,000 km$^2$ and features strong spatial variability in topography and climate.

The format of the manuscript is as follows.  Section~\ref{STmodelSection} introduces the formulation of the double zero-inflated (DZI) spatio-temporal model, details of the model fitting, model assessment, words on inference metrics, and a simulation example, serving as a proof of concept.  Section~\ref{EDA} describes the data used and an exploratory analysis for covariate selection.  Section~\ref{DA} shows the results of application of our proposed modeling to the data described.  Section~\ref{conclusion} summarizes this work and suggests future work to be developed.

\section{Methodology}\label{STmodelSection}
In this section, we specify our DZI model.  We consider the associated distribution theory and model fitting.  We take up model adequacy and model comparison, and we supply useful inference metrics available under the model. A simulation example as a proof of concept is shown.

\subsection{A spatio-temporal model with double zero-inflation and censoring}
The proposed framework adopts a ``double zero-inflation'' approach to distinguish between structural zeros (true dry events) and threshold zeros generated by the detection limit of the measuring device. To this end, we define a latent precipitation process representing the true precipitation mechanism and an observable process subject to left-censoring, whereby precipitation amounts below the detection threshold are recorded as zero.

Let $Y^{\text{true}}_{t\ell}(\mathbf{s})$ and $Y^{\text{obs}}_{t\ell}(\mathbf{s})$ denote the true and observed precipitation, respectively, for day $\ell = 1, \dots, L$, within year $t = 1, \dots, T$, at location $\mathbf{s} \in D \subset \mathbb{R}^2$, where $D$ is the region of interest. If $\epsilon_{t\ell}(\mathbf{s})$ denotes the minimum detection level for day $\ell$ in year $t$ at location $\mathbf{s}$, the resulting relationship can be written as follows:
\begin{equation*}
Y^{\text{obs}}_{t\ell}(\mathbf{s}) = 
\begin{cases}
0, & \text{if } Y^{\text{true}}_{t\ell}(\mathbf{s}) < \epsilon_{t\ell}(\mathbf{s}), \\
Y^{\text{true}}_{t\ell}(\mathbf{s}), & \text{if } Y^{\text{true}}_{t\ell}(\mathbf{s}) \geq \epsilon_{t\ell}(\mathbf{s}).
\end{cases}
\end{equation*}
We assume that the ``latent'' variables $Y^{\text{true}}_{t\ell}(\mathbf{s})$ are 
conditionally independent given the distribution parameters, with their mixture 
distribution specified as:
\begin{equation*}
Y^{\text{true}}_{t\ell}(\mathbf{s}) \sim \pi_{t\ell}(\mathbf{s}) \cdot \delta_0(y) + (1 - \pi_{t\ell}(\mathbf{s})) \cdot f_{Y_{t\ell}(\mathbf{s})}(y), \qquad y \ge 0.
\end{equation*}
Here, $\delta_0(\cdot)$ is a Dirac
delta function at zero, $f_{Y_{t\ell}(\mathbf{s})}(y)$ is a continuous density function on all of $\mathbb{R}^+$, and $0 < \pi_{t\ell}(\mathbf{s}) < 1$. We interpret $\pi_{t\ell}(\mathbf{s})$ as the probability that the conditions for precipitation do not exist for day $\ell$ in year $t$ at location $\mathbf{s}$. 

Furthermore,
\begin{equation} \label{eq:pc1}
\mathbb{P}(Y^{\text{obs}}_{t\ell}(\mathbf{s}) = 0, Y^{\text{true}}_{t\ell}(\mathbf{s}) > 0) = (1 - \pi_{t\ell}(\mathbf{s})) \int_0^{\epsilon_{t\ell}(\mathbf{s})} f_{Y_{t\ell}(\mathbf{s})}(y) \, dy,
\end{equation}
can be interpreted as the probability that the conditions for precipitation exist on day $\ell$ in year $t$ at location $\mathbf{s}$, but precipitation was less than measurable.

The above specification induces the model for $Y^{\text{obs}}_{t\ell}(\mathbf{s})$, a mixed discrete-continuous distribution with a point mass at zero and a continuous positive component left truncated at $\epsilon_{t\ell}(\bs)$. Unlike customary ZI models, we can never observe any data values between $0$ and $\epsilon_{t\ell}(\bs)$.  We can see that under this model, the probability of observing a zero can be written as:
\begin{equation} \label{eq:POZ}
\mathbb{P}(Y^{\text{obs}}_{t\ell}(\mathbf{s}) = 0) = \pi_{t\ell}(\mathbf{s}) + (1 - \pi_{t\ell}(\mathbf{s})) \int_0^{\epsilon_{t\ell}(\mathbf{s})} f_{Y_{t\ell}(\mathbf{s})}(y) \, dy.
\end{equation}
This allows us to separate the two causes of the zero observation introduced by the threshold $\epsilon_{t\ell}(\bs)$.

\subsubsection{Model specification and prior distributions}

Regarding the model specification, we model the probability of a wet event, $1 - \pi_{t\ell}(\bs)$, using a probit link, although a logit link would also provide a reasonable alternative. Specifically, we adopt a linear model for $\eta_{t\ell}(\bs) = \Phi^{-1}(1 - \pi_{t\ell}(\bs))$, where $\Phi(\cdot)$ denotes the standard normal cumulative distribution function (cdf). For positive precipitation amounts, $Y^{\text{true}}_{t\ell}(\bs) \mid Y^{\text{true}}_{t\ell}(\bs) > 0$, and following established practice in precipitation modeling, we adopt a Gamma distribution \citep[see, e.g.,][]{wilks1990}. This choice provides the necessary flexibility to capture precipitation variability, with the distribution characterized by a mean $\mu_{t\ell}(\bs)$ and a global dispersion parameter $\phi$ (inverse of the shape). The probability density function (pdf) of the Gamma distribution, $G(\,\cdot \mid \mu_{t\ell}(\bs), \phi)$, as a member of the exponential family is given by:
\begin{align*}
    f_{Y_{t\ell}(\bs)}(y) =
    \exp\left\{ - \frac{1}{\phi} \left[   \frac{y}{\mu_{t\ell}(\bs)} + \log\left\{\mu_{t\ell}(\bs)\right\} \right] \right\} \, \frac{y^{1/\phi - 1}}{\phi^{1/\phi} \, \Gamma\left(1/\phi\right)}.
\end{align*}
Specifically, we adopt a linear model for $\log \{ \mu_{t\ell}(\bs) \}$. Both linear components for the occurrence and the intensity can include relevant covariates and spatio-temporal random effects, allowing for the comparison of various specifications.

In particular, the linear models are specified as follows:
\begin{equation}\label{model}
    \eta_{t\ell}(\bs) = \mathbf{x}_{t\ell}(\bs) ^{\top}\bm{\beta} + \beta_0(\bs) 
    \quad \text{and} \quad
    \log\{\mu_{t\ell}(\bs)\} = \mathbf{x}_{t\ell}(\bs)^{\top} \bm{\gamma}  + \gamma_0(\bs).
\end{equation}
Here, $\mathbf{x}_{t\ell}(\bs)$ is a $p$-dimensional column vector of covariates (including an intercept) for day $\ell$ of year $t$ at location $\bs$. This vector may include spatial covariates related to geographical or orographic characteristics, temporal covariates associated with seasonality or large-scale climate variability, or spatio-temporal terms, such as a function conditional on the occurrence of precipitation on the previous day, to capture short-term temporal dependence. Although we use the same notation for the occurrence and intensity components, the corresponding covariate vectors may differ in practice. Further, $\bm{\beta}$ and $\bm{\gamma}$ denote $p$-dimensional vectors of regression coefficients. 

To capture spatial dependence, we use spatial Gaussian processes. In particular, $\beta_0(\bs)$ and $\gamma_0(\bs)$ are modeled as independent Gaussian processes with zero mean and exponential covariance functions, parameterized by variance and decay parameters $(\sigma^2_{\beta}, \phi_{\beta})$ and $(\sigma^2_{\gamma}, \phi_{\gamma})$, respectively.\footnote{The covariance function is defined by $C(\bs,\bs';\sigma^2,\phi) = \sigma^2 \exp\{ - \phi \| \bs - \bs' \| \}$, where $\| \bs - \bs' \|$ denotes the Euclidean distance in km between $\bs,\bs' \in D$ computed under the projected coordinate reference system for peninsular Spain, namely Madrid 1870 (Madrid) / Spain LCC (EPSG:2062).}  However, to improve model identifiability, we employ hierarchical centering of the global intercepts and spatial covariates over the Gaussian processes \citep{gelfand1995}, whereby these spatially varying terms provide local adjustments to both the global baselines and the covariate effects. 

For the binary occurrence component, we follow the data augmentation approach of \cite{albert1993}. We expand the parameter space by introducing the latent normal variables $Z_{t\ell}(\bs)$ such that:
\begin{equation*}
Z_{t\ell}(\bs) = \mathbf{x}_{t\ell}(\bs)^{\top} \bm{\beta} + \beta_0(\bs) + \varepsilon_{t\ell}(\bs), \quad \varepsilon_{t\ell}(\bs) \sim \text{i.i.d. } N(0,1).
\end{equation*}
This specification induces the probit model for precipitation occurrence, establishing the link with the true precipitation data such that $Y^{\text{true}}_{t\ell}(\bs) = 0$ if $Z_{t\ell}(\bs) \le 0$, and $Y^{\text{true}}_{t\ell}(\bs) > 0$ if $Z_{t\ell}(\bs) > 0$.

Regarding the prior distributions, we assign conditionally conjugate independent priors across all blocks of parameters when available as follows. Regression coefficients are assigned multivariate normal priors, and we adopt an improper flat prior on the log scale for the Gamma dispersion parameter, i.e.,
\begin{equation} \label{eq:priors1}
\bm{\beta} \sim N_p(\mathbf{a}_{\beta}, \mathbf{B}_{\beta}), 
\quad 
\bm{\gamma} \sim N_p(\mathbf{a}_{\gamma}, \mathbf{B}_{\gamma}),
\quad
[\log(\phi)] \propto 1.
\end{equation}
For the spatial process hyperparameters, we assign independent Gamma priors to the precision parameters and to the decay parameters,
\begin{equation} \label{eq:priors2}
  1/\sigma^2_{\beta}, 1/\sigma^2_{\gamma} \sim G(a_\sigma,b_\sigma), \quad \phi_{\beta}, \phi_{\gamma} \sim G(a_\phi,b_\phi).
\end{equation}
Here, we use the shape-rate parameterization of the Gamma distribution for convenience. For the decay parameters, a prior specification should be considered based on the effective spatial range within the region of interest, defined as $3/\phi_{\beta}$ and $3/\phi_{\gamma}$ for each spatial process, interpreted as the distance beyond which spatial association becomes negligible \citep[][Chapter~2]{banerjee2025}. In particular, this prior gives a mean and variance to the effective ranges of $3 b_\phi / (a_\phi - 1)$ km (for $a_\phi > 1$) and $9 b_\phi^2 / [(a_\phi - 1)^2(a_\phi - 2)]$ km$^2$ (for $a_\phi > 2$), respectively.

\subsubsection{Joint posterior distribution and model fitting}
Let $S = \{\bs_1, \dots,\bs_n\} \subset D$ denote the set of observed precipitation stations. Let $\mathbf{y}^{\text{obs}}$, $\mathbf{Y}^{\text{true}}$, $\mathbf{Z}$, and $\bm{\epsilon}$ denote the vectors collecting all observed values, true precipitation values, latent $Z_{t\ell}(\bs)$ variables, and detection thresholds $\epsilon_{t\ell}(\bs)$, respectively. Denote the spatial processes, $\beta_0(\bs)$ and $\gamma_0(\bs)$, at the observed locations as $\bm{\beta}_0 = (\beta_0(\bs_1), \dots, \beta_0(\bs_n))^\top$ and $\bm{\gamma}_0 = (\gamma_0(\bs_1), \dots, \gamma_0(\bs_n))^\top$, respectively. Further, let $\bm{\theta}$ denote the spatial processes, model parameters, and hyperparameters: for the probit model $\bm{\beta}$ and $\bm{\beta}_0$ with $(\sigma_{\beta}^2, \phi_{\beta})$; for the Gamma model $\bm{\gamma}$, $\phi$, and $\bm{\gamma}_0$ with $(\sigma_{\gamma}^2, \phi_{\gamma})$; and the threshold values $\bm{\epsilon}$. We treat the threshold values as fixed, since external information on the detection thresholds is available. Nevertheless, inference on these quantities could also be incorporated within the proposed framework; see the Appendix~\ref{MCMC} for further details.

Then, the joint posterior distribution of the model, which captures both the occurrence and intensity processes, is proportional to the product of the augmented likelihood and the priors on all spatial processes, model parameters, and hyperparameters:
{\footnotesize
\begin{align*}
    [\bm{\theta}, \mathbf{Z}, \mathbf{Y}^{\text{true}} \mid \mathbf{y}^{\text{obs}}] &\propto 
    [\mathbf{y}^{\text{obs}} \mid \mathbf{Y}^{\text{true}}, \bm{\theta}] \, [\mathbf{Y}^{\text{true}} \mid \mathbf{Z}, \bm{\theta}] \, [\mathbf{Z} \mid \bm{\theta}] \, [\bm{\theta}] \\ &\hspace{-20mm}\propto
    \prod_{i=1}^{n}\prod_{\ell=1}^{L}\prod_{t=1}^{T} \bigg\{ \left( \delta_{Y_{t\ell}^{\text{true}}(\mathbf{s}_i)}(y_{t\ell}^{\text{obs}}(\mathbf{s}_i)) \cdot \mathbf{1}{\{Y_{t\ell}^{\text{true}}(\mathbf{s}_i) \ge \epsilon_{t\ell}(\bs_i)\}} + \delta_{0}(y_{t\ell}^{\text{obs}}(\mathbf{s}_i)) \cdot \mathbf{1}{\{Y_{t\ell}^{\text{true}}(\mathbf{s}_i) < \epsilon_{t\ell}(\bs_i)\}}\right) \\ 
    & \hspace{-10mm} \times \left( \mathbf{1}{\{Z_{t\ell}(\mathbf{s}_i) \le 0\}} \cdot \delta_{0}(Y_{t\ell}^{\text{true}}(\mathbf{s}_i)) + \mathbf{1}{\{Z_{t\ell}(\mathbf{s}_i) > 0\}} \cdot G(Y_{t\ell}^{\text{true}}(\mathbf{s}_i) \mid \mu_{t\ell}(\mathbf{s}_i), \phi) \right) \\ 
    & \hspace{-10mm} \times N(Z_{t\ell}(\mathbf{s}_i) \mid \eta_{t\ell}(\mathbf{s}_i), 1) \bigg\} \times [\bm{\theta}],
\end{align*}
}
where $[\bm{\theta}]$ collects the prior distributions of all spatial processes, model parameters, and hyperparameters. Here, $\delta_{Y_{t\ell}^{\text{true}}(\mathbf{s}_i)}(\cdot)$ is a Dirac delta function at $Y_{t\ell}^{\text{true}}(\mathbf{s}_i)$, $\mathbf{1}{\{\cdot\}}$ is an indicator function, and $G(\,\cdot \mid \mu, \phi)$ and $N(\,\cdot \mid \mu, \sigma^2)$ denote the pdf of a Gamma distribution with mean $\mu$ and dispersion $\phi$ and a normal distribution with mean $\mu$ and variance $\sigma^2$, respectively.

Posterior inference is conducted within a Bayesian framework using MCMC methods; full distributional details, including the treatment of missing data, are provided in the Appendix~\ref{MCMC}. The outline of the proposed Metropolis-within-Gibbs algorithm consists of the following steps:
\begin{enumerate}
    \item There exists structural dependence between $Z_{t\ell}(\bs_i)$ and $Y^{\text{true}}_{t\ell}(\bs_i)$, so that the Markov chain cannot move between regimes under separate updates. Therefore, we jointly update the pair from their full conditional distribution:
    \begin{equation*}
      [\mathbf{Z}, \mathbf{Y}^{\text{true}} \mid \mathbf{y}^{\text{obs}}, \bm{\theta}] = [\mathbf{Y}^{\text{true}} \mid \mathbf{y}^{\text{obs}}, \bm{\theta}] \, [\mathbf{Z} \mid \mathbf{Y}^{\text{true}}, \bm{\theta}].
    \end{equation*}
    This involves sampling $Y^{\text{true}}_{t\ell}(\bs_i)$ from a mixture of a point mass at zero and a Gamma distribution right-truncated at $\epsilon_{t\ell}(\bs_i)$, followed by sampling $Z_{t\ell}(\bs_i)$ from a truncated normal distribution conditional on the precipitation state.

    \item There are no closed-form full conditional distributions for the spatial process, the regression coefficients, or the dispersion parameter of the Gamma model. We adopt the iterative weighted least squares within a Metropolis-Hastings step with a multivariate normal proposal as suggested by \cite{gamerman1997}, separately for $\bm\gamma_0$ and $\bm{\gamma}$. We update $\log(\phi)$ using an adaptive random-walk Metropolis step, tuning the proposal variance following \cite{roberts2009} to target an acceptance rate of approximately 33\%.

    \item The augmentation of the parameter space proposed by \cite{albert1993} leads to conjugate multivariate normal full conditional distributions for the spatial process and the probit regression coefficients, $\bm{\beta}_0$ and $\bm{\beta}$, respectively.
    
    \item We implemented hierarchical centering of the global intercept and spatial covariates with respect to the spatial processes to improve the mixing of the algorithm \citep{gelfand1995}. Then, the hyperparameters of the spatial processes have a closed-form full conditional distribution available for the mean coefficients and the variance parameter. For the decay parameter, we use a random-walk Metropolis step on the log scale \citep{roberts2009}, with the proposal variance adaptively tuned to a 33\% acceptance rate.
\end{enumerate}

\subsection{Model adequacy and comparison}\label{model.comp}
Model comparison is performed using various cross-validation (CV) metrics and deviance to assess goodness of fit. Furthermore, model adequacy is examined via posterior predictive checks targeting specific features of interest. All diagnostics are evaluated relative to observed precipitation, acknowledging that latent true precipitation occurrences remain inherently unobservable.

To assess the predictive and estimation power of the models, we employ $K$-fold CV. 
The performance evaluation of each fold is divided into two stages. The first stage consists of evaluating the ability of the model to classify the observed zeros using the area under the receiver operating characteristic curve \citep[AUC, ][]{robin2011} and Tjur's $R^2$ \citep{tjur2009}. The latter metric is calculated as the difference between the mean predicted probabilities for wet and dry days. Higher values of AUC and Tjur's $R^{2}$ indicate superior predictive performance, with values closer to their upper bounds of $1$ reflecting sharper classification and group separation, respectively. The second stage assesses the predictive performance of positive values. We employ the mean squared error (MSE) and the continuous ranked probability score \citep[CRPS,][]{gneiting2007} in a Bayesian spatio-temporal framework. In the context of our DZI model, replicates of each positive observation are sampled from a truncated Gamma distribution on the corresponding detection level of the observed value. Superior predictive performance is indicated by lower values of both MSE and CRPS, both bounded below by $0$, reflecting minimized prediction errors and enhanced probabilistic calibration, respectively. Details on the calculation of these metrics may be found in Section S1.1 of the Supplementary material (hereinafter, `S' denotes Supplementary material). 

Model comparison is also considered in terms of deviance. One of the most robust metrics for this purpose is the widely applicable information criterion \citep[WAIC,][Chapter~7]{gelman2013}. WAIC is a fully Bayesian criterion, particularly useful in hierarchical modeling. Unlike the commonly used deviance information criterion (DIC), which relies on a point estimate, WAIC is based on the entire posterior distribution. It is calculated as the sum of two components: the log pointwise predictive density, which measures the model's fit to the observed data, and a penalty term that accounts for the effective number of parameters to prevent overfitting; see Section S1.2. 

Model adequacy or model checking for positive values is analyzed in terms of calibration and sharpness of the probabilistic predictions. One of the most effective tools for assessing calibration is the probability plot (PP-plot). It compares the posterior predictive probabilities against the observed empirical probabilities. A well-calibrated model displays points closely aligned with the unit diagonal. This approach is intrinsically linked to the probability integral transform (PIT), which is the value that the posterior cdf attains at an observation. Under the assumption of a continuous predictive distribution, a correctly specified model results in PIT values following a standard uniform distribution \citep{angus1994}. Section S1.3 includes all the details for the computation of these tools. Additionally, quantile plots (QQ-plots) are used, where we compare the empirical observed quantiles and the expected quantiles under the model specification. The PP-plot and QQ-plot contain the same information but expressed on different scales.

\subsection{Model-based inference metrics for missing precipitation}\label{inference_met}
One of our inference objectives is the posterior predictive distribution of the true latent precipitation process across any location within the study region and period, i.e., $Y^{\text{true}}_{t\ell}(\bs_0)$ for any day $\ell = 1, \dots, L$, within year $t = 1, \dots, T$, and location $\bs_0 \in D$. From the linear models in Equation~\ref{model}, a sample of $Y^{\text{true}}_{t\ell}(\bs_0)$ can be obtained by first sampling from a Bernoulli distribution with probability $1 - \pi_{t\ell}(\bs_0)$, which is the probability of a wet event. A sample of $1 - \pi_{t\ell}(\bs_0)$ is obtained by evaluating the standard normal cdf $\Phi(\cdot)$ at the linear predictor $\eta_{t\ell}(\bs_0)$. In the case of a success from the Bernoulli distribution, a sample of $Y^{\text{true}}_{t\ell}(\bs_0)$ is finally obtained as a sample from a Gamma distribution with mean $\mu_{t\ell}(\bs_0)$ and dispersion parameter $\phi$. If the Bernoulli sample is equal to zero, the sample for $Y^{\text{true}}_{t\ell}(\bs_0)$ is equal to zero, i.e., a dry event. A sample for $Y_{t\ell}^{\text{obs}}(\bs_0)$ is obtained by left-censoring $Y^{\text{true}}_{t\ell}(\bs_0)$ with respect to the detection threshold $\epsilon_{t\ell}(\bs_0)$.
Posterior samples for the parameters are available for each MCMC realization, and posterior samples for the spatial processes $\beta_0(\bs)$ and $\gamma_0(\bs)$ are obtained using the posterior samples of the parameters through Bayesian kriging \citep[][Chapter~6]{banerjee2025}. 

Using $G_D$, a fine spatial grid for $D$, posterior samples of $Y_{t\ell}^\text{true}(\bs_j)$, $\pi_{t\ell}(\bs_j)$, and $\mu_{t\ell}(\bs_j)$, can be computed at each location $\bs_j \in G_D$, for every day $\ell = 1, \dots, L$, within year $t = 1, \dots, T$. This enables us to offer a collection of metrics that provide the inference we seek to extract from the generic DZI spatio-temporal precipitation model; see a summary of these metrics in Table~\ref{table_metrics}.

\begin{table}[]
\scriptsize
    \centering
    \begin{tabular}{l|l}
    \hline
    \textbf{Metric} & \textbf{Description} \\
    \hline 
        PC &  Probability of censoring when conditions for precipitation exist\\
        RZC & Ratio between probability of true zero and probability of censoring \\
        PCD & Percentage of censored days when the observed value is zero\\
        EOPP & Expected value of observed positive precipitation\\
        EUP & Expected undetected precipitation \\
        PRDM & Percentage difference of true precipitation relative to observed precipitation of the mean\\
        PRDQ90 & Percentage difference of true precipitation relative to observed precipitation of the $0.90$ quantile\\
        \hline
    \end{tabular}
    \caption{Name and brief description of the model-based inference metrics.}
    \label{table_metrics}
\end{table}

We first consider the probability of censoring (PC) when precipitation conditions exist. This was given in Equation~\ref{eq:pc1} and repeated below for notational completeness:
\begin{align*}
    \text{PC}_{t\ell}(\bs_j) = \mathbb{P}(Y^{\text{obs}}_{t\ell}(\mathbf{s}_j) = 0, Y^{\text{true}}_{t\ell}(\mathbf{s}_j) > 0) = ( 1 - \pi_{t\ell}(\mathbf{s}_j)) \int_0^\epsilon f_{Y_{t\ell}^\text{true}(\mathbf{s}_j)}(y)\,dy, 
\end{align*}
where $\epsilon$ is the detection level, which in $G_D$, without loss of generality, will remain constant in space and time. Using the same constant threshold, the probability of observing a zero, $\mathbb{P}(Y^{\text{obs}}_{t\ell}(\mathbf{s}) = 0)$, is obtained from Equation~\ref{eq:POZ}.

Closely related to these probabilities, we can calculate the ratio between the probability of a true zero and the probability of censoring (RZC), and the percentage of censored days if the observed value is zero (PCD). These probabilities are defined as:
\begin{align*}
    \text{RZC}_{t\ell}(\mathbf{s}_j) = \frac{\pi_{t\ell}(\mathbf{s}_j)}{PC_{t\ell}(\bs_j)}, \quad
     \text{PCD}_{t\ell}(\bs_j)= \frac{\text{PC}_{t\ell}(\mathbf{s}_j)}{\mathbb{P}(Y^{\text{obs}}_{t\ell}(\mathbf{s}_j) = 0)} \cdot 100\%. 
\end{align*}
The ratio RZC may be interpreted as the number of days with true zeros for each day with censoring, whereas averaging PCD across years and days can be interpreted as the percentage of censored days out of all days with observed zeros.

Additionally, we calculate the expected value of observed positive precipitation (EOPP), as well as the expected value of undetected precipitation (EUP). After some calculations (see Sections S2.1 and S2.2), the final expression for these metrics are:
\begin{align*}
    \text{EOPP}_{t\ell}(\bs_j) = \mu_{t\ell}(\mathbf{s}_j)  \int_\epsilon^\infty f_{\tilde{Y}_{t\ell}(\mathbf{s}_j)}(y) \, dy, \quad \text{EUP}_{t\ell}(\bs_j) = ( 1 - \pi_{t\ell}(\mathbf{s}_j)) \mu_{t\ell}(\mathbf{s}_j) \int_0^\epsilon f_{\tilde{Y}_{t\ell}(\mathbf{s}_j)} (y) \, dy, 
\end{align*}
where $\tilde{Y}_{t\ell}(\mathbf{s}_j) \sim G(1 + 1 / \phi, \, 1 / (\mu_{t\ell}(\mathbf{s}_j) \phi) )$ (here, shape-rate parameterization for convenience).

Lastly, we can also compare the mean and $0.90$ quantile of observed and true precipitation. We do this by calculating the difference between the EOPP and the expected true precipitation $\mu_{t\ell}(\bs)$ expressed as a percentage relative to the EOPP, and the difference between the observed and true precipitation conditional quantiles on positive values expressed as a percentage of the observed conditional quantile, i.e.,
\begin{align*}
    \text{PRDM}_{t\ell}(\bs_j) &= \left(1 -\frac{\mu_{t\ell}(\bs_j)}{\text{EOPP}_{t\ell}(\bs_j)}\right) \cdot 100\%, \quad \\ \text{PRDQ90}_{t\ell}(\bs_j) &= \left( 1 - \frac{Q_{0.90}(Y_{t\ell}^\text{true}(\bs_j) \mid Y_{t\ell}^\text{true}(\bs_j) > 0)}{Q_{0.90}(Y_{t\ell}^\text{obs}(\bs_j) \mid Y_{t\ell}^\text{obs}(\bs_j) > \epsilon)} \right) \cdot 100\%, 
\end{align*}
where $Q_{0.90}(\,\cdot \mid \cdot\,)$ is the $0.90$ conditional quantile. The observed precipitation quantile corresponds to a quantile of a truncated Gamma distribution; see Section S2.3. 

To enrich the framework, one could introduce a \emph{conditional} binary predictor into the linear models in Equation~\ref{model}, i.e., $I_{t \ell}(\bs)$ indicating whether precipitation that exceeded $1$ mm was observed on the previous day ($I_{t\ell}(\bs)=1$ if $Y^{\text{obs}}_{t,\ell-1}(\bs) > 1$) or not ($I_{t\ell}(\bs) =0$ if $Y^{\text{obs}}_{t,\ell-1}(\bs) \le 1$). In line with WMO-recommended precipitation criteria \citep{kleintank2009}, the 1 mm threshold is utilized to capture the temporal persistence of regional precipitation, effectively partitioning the previous day predictor into events below 1 mm, including zeros, and higher-intensity regimes.

This leads to three versions of $f_{Y_{t\ell}^\text{true}(\bs)}(y)$ above: two conditionals, $f_{Y_{t\ell}^\text{true}(\bs) \mid I_{t\ell}(\bs)=1}(y)$ and $f_{Y_{t\ell}^\text{true}(\bs) \mid I_{t\ell}(\bs)=0}(y)$ as well as the marginal, $f_{Y_{t\ell}^\text{true}(\bs)}(y)$.  We explore each of these in our ensuing inference. 

\subsection{A simulation example}\label{simulation_ex}
This simulation aims to demonstrate the capability of the DZI models to recover model parameters, and correctly capture true zeros in data censored at a specific detection level, given that the exact number of censored data points is known by design. For this reason, we fit the simulated data using two different approaches. First, as a naive baseline, we assume the data is uncensored, applying a standard ZI approach despite the presence of actual censoring. Secondly, we employ our proposed DZI approach. To evaluate the performance of both approaches, we compare the estimated coefficients against the true parameter values used to generate the simulated data.

We simulate $1000$ synthetic datasets each with $N = 10{,}000$ observations (organized in $100$ days within $10$ years in $10$ different stations) in the following simulation examples (SE) of the model in Equation~\ref{model}:
\begin{align*}
    (\text{SE1}) \quad  \eta_{t\ell}(\bs) &= 0.1, &\log\{\mu_{t\ell}(\bs) \} &= 1.35, &\phi &= 3; \\
    (\text{SE2}) \quad \eta_{t\ell}(\bs) &= -0.3 + 0.9 \cdot I_{t\ell}(\bs), &\log\{\mu_{t\ell}(\bs) \} &= 1.2 + 0.4 \cdot I_{t\ell}(\bs), &\phi &= 3,
\end{align*}
where $\phi$ is the dispersion parameter of the Gamma distribution, and $I_{t\ell}(\bs) = \mathbf{1}\{Y_{t,\ell-1}^{\text{obs}}(\bs) > 1\}$ is the conditional binary variable. The detection level is fixed to $\epsilon = 0.1$ for all days and locations, i.e., simulated data below this value were censored to zero. We selected coefficient values to match observed patterns: approximately $55$--$65\%$ zeros, a mean nonzero intensity of $5$--$6$ mm, and a temporal persistence in SE2 close to that of the observed precipitation data.

The models were fitted within a Bayesian framework using $10{,}000$ burn-in iterations followed by $10{,}000$ sampling iterations; applying a thinning interval of $10$ yielded a final pool of $1000$ retained samples per chain.

Table~\ref{sim_table} shows the bias, $95\%$ credible band coverage (CVG), and root mean square error (RMSE) averaged across the $1000$ synthetic datasets. For each simulation example, these metrics are displayed for all model parameters on the left, and for the model-based inference metrics, PC and EUP, on the right. Coefficients of the probit model are denoted as $\beta_0$ (intercept) and $\beta_1$ (indicator coefficient), whereas in the Gamma model they are denoted as $\gamma_0$ and $\gamma_1$, respectively. 

The DZI model provides very satisfactory results for all metrics, successfully capturing the amount of missing precipitation produced by left-censoring as well as recovering the data simulation coefficients. By contrast, a standard ZI model yields poor results in recovering model parameters, highlighting the advantages of the double zero approach. Because standard ZI models are inherently unequipped to recover the latent true precipitation, their parameter estimates exhibit severe bias, which systematically dominates the RMSE, and their empirical CVG drop to practically zero in all cases, failing to capture the true generating values. 

\begin{table}[]
    \centering
    \footnotesize 
    \setlength{\tabcolsep}{3.5pt} 

    \begin{minipage}[b]{0.51\textwidth}
        \centering
        \begin{tabular}{l|c|cc|cc}
        \hline
         \multicolumn{2}{c|}{\multirow{2}{*}{\textbf{Sim. Example}}} & \multicolumn{2}{c|}{\textbf{SE1}} & \multicolumn{2}{c}{\textbf{SE2}} \\
         \cmidrule(lr){3-4} \cmidrule(lr){5-6}
         \multicolumn{2}{c|}{} & ZI & DZI & ZI & DZI \\
         \hline
         \multirow{3}{*}{$\gamma_0$} & Bias & $0.259$ & $-0.002$ & $0.274$ & $-0.004$ \\
         & CVG & $0$ & $0.941$ & $0$ & $0.939$  \\
         & RMSE & $0.260$ & $0.043$ & $0.275$ & $0.056$ \\
        \hline 
        \multirow{3}{*}{$\gamma_1$} 
         & Bias &  &  & $-0.038$ & $0$ \\
         & CVG &  &  & $0$ & $0.951$ \\
         & RMSE &  &  & $0.071$ & $0.076$ \\
        \hline 
        \multirow{3}{*}{$\beta_0$} 
         & Bias & $-0.311$ & $0.005$ & $-0.253$ & $0.005$ \\
         & CVG & $0$ & $0.949$ & $0$ & $0.930$ \\
         & RMSE & $0.311$ & $0.043$ & $0.253$ & $0.041$ \\
        \hline 
        \multirow{3}{*}{$\beta_1$} 
         & Bias &  &  & $-0.163$ & $0.010$ \\
         & CVG &  &  & $0.801$ & $0.936$ \\
         & RMSE &  &  & $0.166$ & $0.070$ \\
        \hline 
        \multirow{3}{*}{$\phi$} 
         & Bias & $-1.495$ & $0.018$ & $-1.494$ & $0.035$ \\
         & CVG & $0$ & $0.952$ & $0$ & $0.935$  \\
         & RMSE & $1.495$ & $0.204$ & $1.495$ & $0.228$  \\
        \hline 
        \end{tabular}
    \end{minipage}
    \hfill 
    \begin{minipage}[b]{0.45\textwidth}
        \centering
        \begin{tabular}{l|c|c|cc}
        \hline
         \multicolumn{2}{c|}{\multirow{2}{*}{\textbf{Sim. Example}}} & \multicolumn{1}{c|}{\multirow{2}{*}{\textbf{SE1}}} & \multicolumn{2}{c}{\textbf{SE2}} \\
          \cline{4-5}
         \multicolumn{2}{c|}{} &  & $I_{t\ell}(\bs) = 0$ & $I_{t\ell}(\bs) = 1$  \\
         \hline
         \multirow{3}{*}{PC} & Bias & $0.002$ & $0.002$ & $0.004$  \\
         & CVG & $0.949$ & $0.928$ & $0.935$   \\
         & RMSE & $0.015$ & $0.012$ & $0.022$  \\
        \hline 
        \multirow{3}{*}{EUP} 
         & Bias & $0$ & $0$ & $0$  \\
         & CVG & $0.950$ & $0.923$ & $0.933$  \\
         & RMSE & $0$ & $0$ & $0$ \\
        \hline 
        \end{tabular}
    \end{minipage}

    \vspace{3mm} 
    \begin{minipage}[t]{0.51\textwidth}
        \small \textbf{(a)} Bias, CVG and RMSE values for the parameters of the simulation examples  SE1 and SE2, for ZI and DZI model fitting techniques.
        \label{sim_table1}
    \end{minipage}
    \hfill
    \begin{minipage}[t]{0.45\textwidth}
        \small \textbf{(b)} Bias, CVG and RMSE values of PC and EUP of the simulation examples SE1 and SE2, for the DZI model fitting technique.
        \label{sim_table2}
    \end{minipage}

    \normalsize 
    \caption{Results for the simulation examples. Values displayed as zero are due to rounding to three decimal places.}
    \label{sim_table}
\end{table}

\section{The dataset and covariate selection}\label{EDA}

Here, we describe the dataset in detail including the thresholding incorporated in the data as well as some empirical precipitation summaries.  We then turn to variable selection using a probit model for explanation of incidence and a hurdle model for explanation of positive precipitation amounts.

\subsection{Description of the dataset} 
\begin{figure}[t]
    \centering
    \includegraphics[width = 0.8\linewidth]{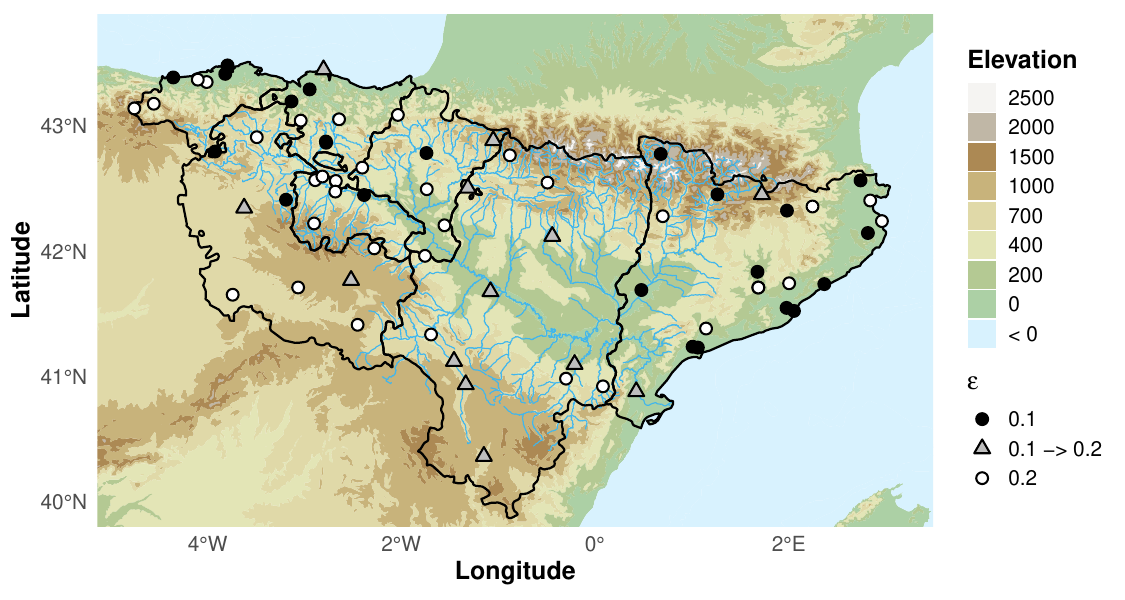}
    \caption{Map of the 70 observed locations and their detection thresholds during the study period (MAM 2010--2024), together with the region of interest by autonomous community.}
    \label{fig:map}
\end{figure}

We used data from automatic stations equipped with tipping-bucket rain gauges \citep[][Chapter~6]{WMO2024}, which record liquid precipitation amounts in millimeters (mm), where 1 mm corresponds to 1 liter per square meter. The point-referenced dataset consists of $70$ daily precipitation series from AEMET (the Spanish Meteorological Office) located around the Ebro River Basin, spanning the autonomous communities of Aragon, Basque Country, Cantabria, Catalonia, Navarre, La Rioja, and the provinces of Burgos and Soria in Castile and León (see Figure~\ref{fig:map}). These regions cover most of the Ebro River Basin, which constitutes our region of interest. The maximum distance between stations is of approximately $670$ km. The time series include daily observations from March to May (MAM), corresponding to the spring season, over the period 2010--2024. All retained series have data availability exceeding $80$\% during the study period, with $46$ series exceeding $95$\%. Overall, the dataset contains $70 \text{ (sites)} \times 15 \text{ (years)} \times 92 \text{ (days)} = 96{,}600$ potential observations, of which 4588 are missing values, corresponding to less than 5\% of the data.

The Ebro River Basin is located in the northeastern part of Spain and has an area of approximately 85,000 km$^2$. The maximum elevation is roughly 3400 m in the Pyrenees, 2600 m in the Iberian System, and between 200 and 400 m in the Central Valley. Most of the area is characterized by a Mediterranean-Continental dry climate with irregular precipitation and a large temperature range. However, climate differences can be distinguished by elevation and the influence from the Mediterranean Sea in the east and the Cantabrian Sea in the north \citep{aemet11}. These factors result in pronounced spatial heterogeneity across the basin, with strong gradients in both topography and climate.

\subsection{Minimum detection threshold by site and year}
Tipping-bucket rain gauges record precipitation in discrete increments corresponding to a fixed bucket volume, typically 0.1 or 0.2 mm, so that precipitation amounts below this detection threshold are not registered and appear as zeros \citep[][Chapter~6]{WMO2024}. This mechanism induces left-censoring, making small but positive precipitation amounts indistinguishable from true dry events. In practice, the effective detection threshold may vary across sites and over time due to differences in instrumentation, calibration, or recording protocols; as above,  for year $t$, day $\ell$ and site $\bs$, we denote this threshold as $\epsilon_{t\ell}(\bs)$. Although the threshold formally truncates the positive precipitation distribution, its practical impact is largely concentrated at zero. Consequently, our effort focuses on accurately characterizing the occurrence process and the censoring-induced excess of zeros.

We first developed an exploratory analysis for the detection thresholds at each station across years. Detection thresholds vary both spatially and temporally, taking values of either 0.1 or 0.2 mm. In particular, $24$ and $33$ out of $70$ stations maintain thresholds equal to $0.1$ and $0.2$ mm across the study period, respectively, whereas $13$ stations transition from 0.1 to 0.2 mm at a specific point in time. Figure~\ref{fig:map} shows a spatial summary of the detection thresholds across the whole period of study by station. Section~S3.1 provides details on the selection of the thresholds and their corresponding values for each year and location.

\subsection{Empirical description of precipitation}
Here, we provide a summary of the spatial patterns of precipitation occurrence and intensity over the study period. We then explore long-term temporal relationships with large-scale climate indices, as well as short-term temporal dependence.

\begin{figure}[t]
    \centering
    \includegraphics[width=0.47\linewidth]{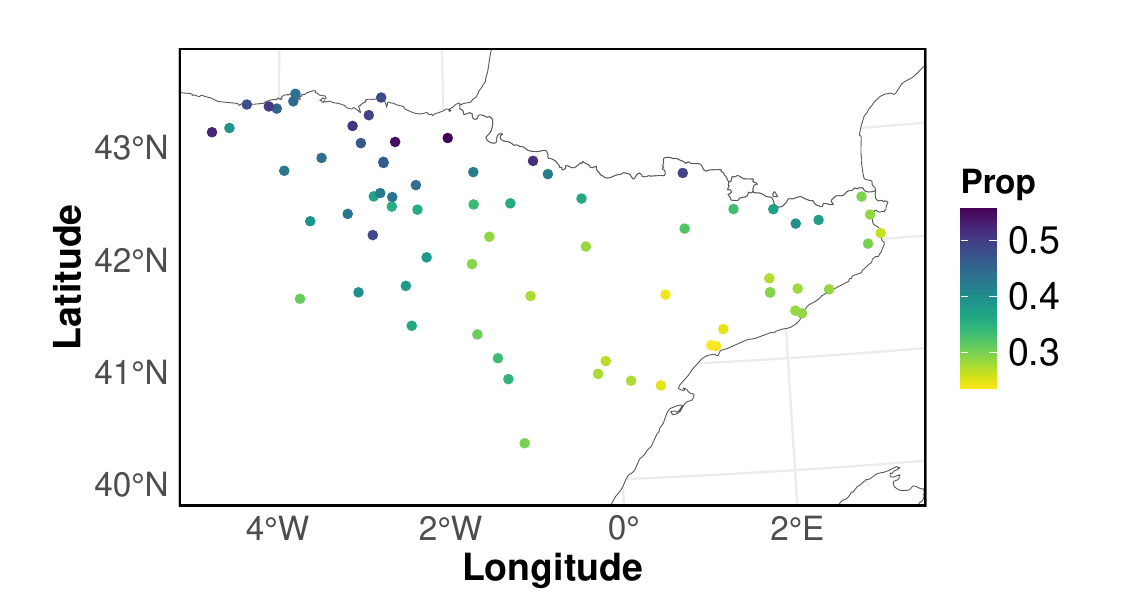}
    \includegraphics[width=0.47\linewidth]{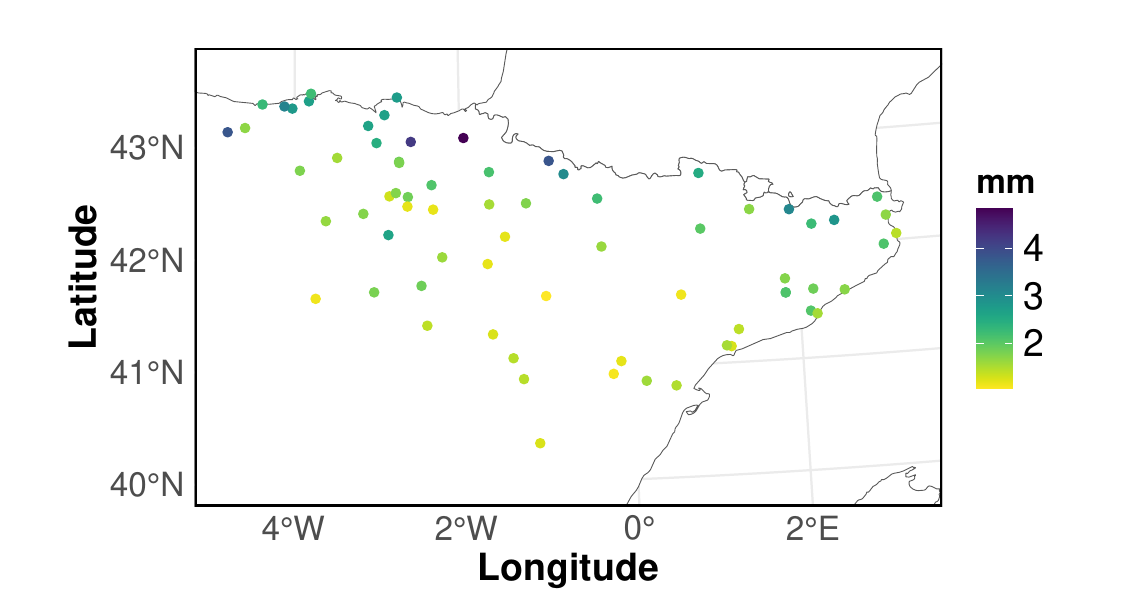}
    \caption{Left: Proportion of wet days. Right: Mean precipitation amount conditional on positive values. MAM 2010--2024.}
    \label{fig:prop_0}
\end{figure}

Figure~\ref{fig:prop_0} shows the proportion of wet days (left plot) and the mean precipitation amount conditional on positive values (right plot) in the whole study period. Both maps reveal clear spatial dependence; lower precipitation frequency is observed along the Mediterranean coast, whereas higher occurrence and intensity are found in the Cantabrian coast and the Pyrenees. 

Annual variability reveals clear temporal pattern, with alternating wet and dry years throughout the study period. Different variables derived from the monthly North Atlantic Oscillation (NAO) index, obtained from the National Centers for Environmental Information dataset, were evaluated to explain large-scale temporal precipitation behavior. We considered the monthly value series and their lags up to $3$ months, as well as the 3-month moving averages and their lags up to $12$ months \citep{trigo2004}. The mean value from August to October of the previous year exhibits the strongest relationship with both precipitation occurrence and intensity. Specifically, an increase in this NAO variable yields an decreasing trend in the proportion of wet days and a decreasing trend in the mean precipitation of wet days; see Section~S3.2. 

Furthermore, we explore temporal dependence at the daily scale by examining precipitation occurrence and intensity conditional on the previous day's events (see Section~S3.3), finding substantial persistence, short-term temporal dependence. This motivates the inclusion of a term in both model components in the form of a conditional binary indicator of previous day precipitation exceeding 1 mm.

\subsection{Exploratory probit and hurdle models for occurrence and intensity}

Next, we identify which available spatial and temporal covariates explain daily precipitation. To facilitate comparison, in this section we employ exploratory generalized linear models (GLM) estimated via maximum likelihood, first for the occurrence and subsequently for the intensity.

We first focus on the occurrence of zero observations, modeling them using probit models. We evaluate a variety of fixed effect combinations, using the AUC, Tjur's $R^2$, and AIC. The fixed effects we employ consist of large-scale and long-term variables (NAO index), short time memory variables (previous day indicator), and geographical variables that describe the relief (elevation) and location (log-plus-one distances to both coasts).
The base model includes only elevation, while subsequent models extend this baseline. First, we separately add three components: the log-plus-one distances to both the Cantabrian and Mediterranean coasts, the NAO index, and a conditional binary indicator for whether the previous day's precipitation exceeded 1 mm. This yields three additional models. Lastly, we fit the full model containing all previously described covariates. 
Table~\ref{table_probit_hurdle} shows the results, highlighting the use of the temporal and spatial covariates altogether, especially the previous day precipitation indicator.

\begin{table}[]
    \centering
    \begin{tabular}{l|ccc|ccc}
    \hline
     \multicolumn{1}{c|}{\multirow{2}{*}{\textbf{Model / Metric}}} & \multicolumn{3}{c|}{\textbf{Occurrence}} & \multicolumn{3}{c}{\textbf{Intensity}} \\
     \cmidrule(lr){2-4} \cmidrule(lr){5-7}
     \multicolumn{1}{c|}{} & AUC & Tjur's $R^2$ & AIC & CRPS & MSE & AIC \\
     \hline
     Base model (Elevation) & $0.514$ & $0.001$ & $122480$ & $3.419$ & $64.114$ & $177327$ \\
     $\hspace{.25cm} + \, \log \text{Distances}$ & $0.587$ & $0.021$ & $120765$ & $3.408$ & $63.715$ & $176942$ \\
     $\hspace{.25cm} + \, \text{NAO}$ & $0.541$ & $0.005$ & $122125$ & $3.414$ & $63.981$ & $177230$ \\
     $\hspace{.25cm} + \, \text{Previous day}$ & $0.655$ & $0.120$ & $111929$ & $3.385$ & $63.090$ & $176525$ \\
     Full model & $\mathbf{0.690}$ & $\mathbf{0.131}$ & $\mathbf{110799}$ & $\mathbf{3.370}$ & $\mathbf{62.511}$ & $\mathbf{176078}$ \\
    \hline 
    \end{tabular}
    \caption{In-sample model comparison metrics for the exploratory probit (occurrence) and Gamma (intensity) GLMs. The best results for each metric are marked in bold.}
    \label{table_probit_hurdle}
\end{table}

Secondly, for variable selection regarding positive precipitation, we fit exploratory Gamma models within a hurdle model framework. We compare the same set of fixed effect combinations making use of the CRPS, MSE, and AIC. As in the probit models, the results highlight the use of all studied covariates, with the most remarkable improvement when adding the previous day precipitation indicator; see Table~\ref{table_probit_hurdle}.

\section{Analysis of the data}\label{DA}

This section presents the results of applying the methodology proposed in Section~\ref{STmodelSection} to the dataset described in Section~\ref{EDA}.

All models were fitted by scaling the covariates to have zero mean and unit variance to enhance MCMC mixing. Regarding the prior specification, let $\mathbf{0}_n$ denote the $n$-dimensional vector of zeros, and let $\mathbf{I}_n$ denote the $n \times n$ identity matrix. We assign weakly informative priors for the regression coefficients with hyperparameters in Equation~\ref{eq:priors1} given by $\mathbf{a}_{\beta} = \mathbf{a}_{\gamma} = \mathbf{0}_p$ and $\mathbf{B}_{\beta} = \mathbf{B}_{\gamma} = 10^{6} \, \mathbf{I}_p$. For the spatial process hyperparameters, we consider more informative priors to address the well-known identifiability issues associated with them \citep{zhang2004}. In particular, in Equation~\ref{eq:priors2} we give $a_\sigma = 2$, $b_\sigma = 1$, $a_\phi = 32$, $b_\phi = 2300$. Our specification yields a prior for the effective ranges with a prior mean of about $222$ km, corresponding to approximately one third of the maximum distance in the region of study, and a standard deviation of $40$ km; i.e., it places $95$\% of the probability mass between $157$ and $315$ km.

Model fitting was performed by running two chains of the MCMC algorithm for each model, each starting with distinct initial values and running for 200,000 iterations to collect samples from the joint posterior distribution. The first half of the samples were discarded as burn-in, and the remaining samples were thinned to 500 from each chain. Convergence diagnostics for the MCMC samples of the selected model are provided in Section~S4.1.

\subsection{Model selection}
For model selection, we perform a 10-fold CV and study the validation metrics introduced in Section~\ref{model.comp}, which enable us to evaluate the predictive performance for both occurrence and intensity of precipitation. Each fold is trained in a dataset containing observations from $63$ stations, and metrics are evaluated in the remaining $7$ stations and averaged across folds. Additionally, we calculate the value of the WAIC for the proposed models fitted using the complete dataset. 

We propose the comparison of four different DZI-GLM models, following the expressions in Equation~\ref{model}. All models include a global intercept, the NAO index, elevation, and the log-plus-one distances to both coasts in both linear terms. The first comparison concerns whether the linear predictor is conditioned on the previous day's precipitation; models incorporating this conditional variable are denoted with a `C'. Specifically, we include an indicator for previous day precipitation exceeding $1$ mm as a fixed effect covariate. This results in models with either $p = 6$ or $p = 5$ regression coefficients, depending on whether they include the conditional term. Additionally, we compare models with and without spatial Gaussian processes, denoting those that include these random effects with an `S'. Consequently, these models are labeled as DZI-GLM, DZI-CGLM, DZI-SGLM, and DZI-SCGLM, respectively.

Table~\ref{CV.table} shows the CV results for these models. Although both temporal dependence, via the previous day's precipitation indicator, and spatial random effects improve overall predictive performance, the binary temporal variable shows a stronger influence on occurrence, whereas both terms yield comparable improvements in precipitation intensity. The results indicate that the most complex model, the DZI-SCGLM, performs best for our data in terms of occurrence prediction, intensity prediction, and deviance. 

\begin{table}[t]
\centering
\begin{tabular}{l | cc | cc | c}
  \hline
  \multicolumn{1}{c|}{\multirow{2}{*}{\textbf{Model / Metric }}} & \multicolumn{2}{c|}{\textbf{Occurrence}} & \multicolumn{2}{c|}{\textbf{Intensity}} & \multicolumn{1}{c}{\textbf{Deviance}}\\
  \cmidrule(lr){2-3} \cmidrule(lr){4-5} \cmidrule(lr){6-6}
    & AUC & Tjur's $R^2$ & CRPS & MSE & WAIC \\
  \hline
  DZI-GLM & $0.587$ & $0.023$ & $3.40$ & $65.1$ & $283810$\\
  DZI-CGLM & $0.690$ & $0.115$ & $3.37$ & $64.1$ & $274106$ \\
   DZI-SGLM & $0.591$ & $0.026$ & $3.38$ & $64.4$ & $282418$\\
   DZI-SCGLM & $\mathbf{0.692}$ & $\mathbf{0.116}$ & $\mathbf{3.35}$ & $\mathbf{63.4}$ & $\mathbf{273241}$\\
   \hline
\end{tabular}
\caption{10-fold CV model comparison metrics for the DZI-GLM models. The best results for each metric are marked in bold.}
\label{CV.table}
\end{table}

\begin{figure}[tb]
    \centering
    \includegraphics[width=0.32\linewidth]{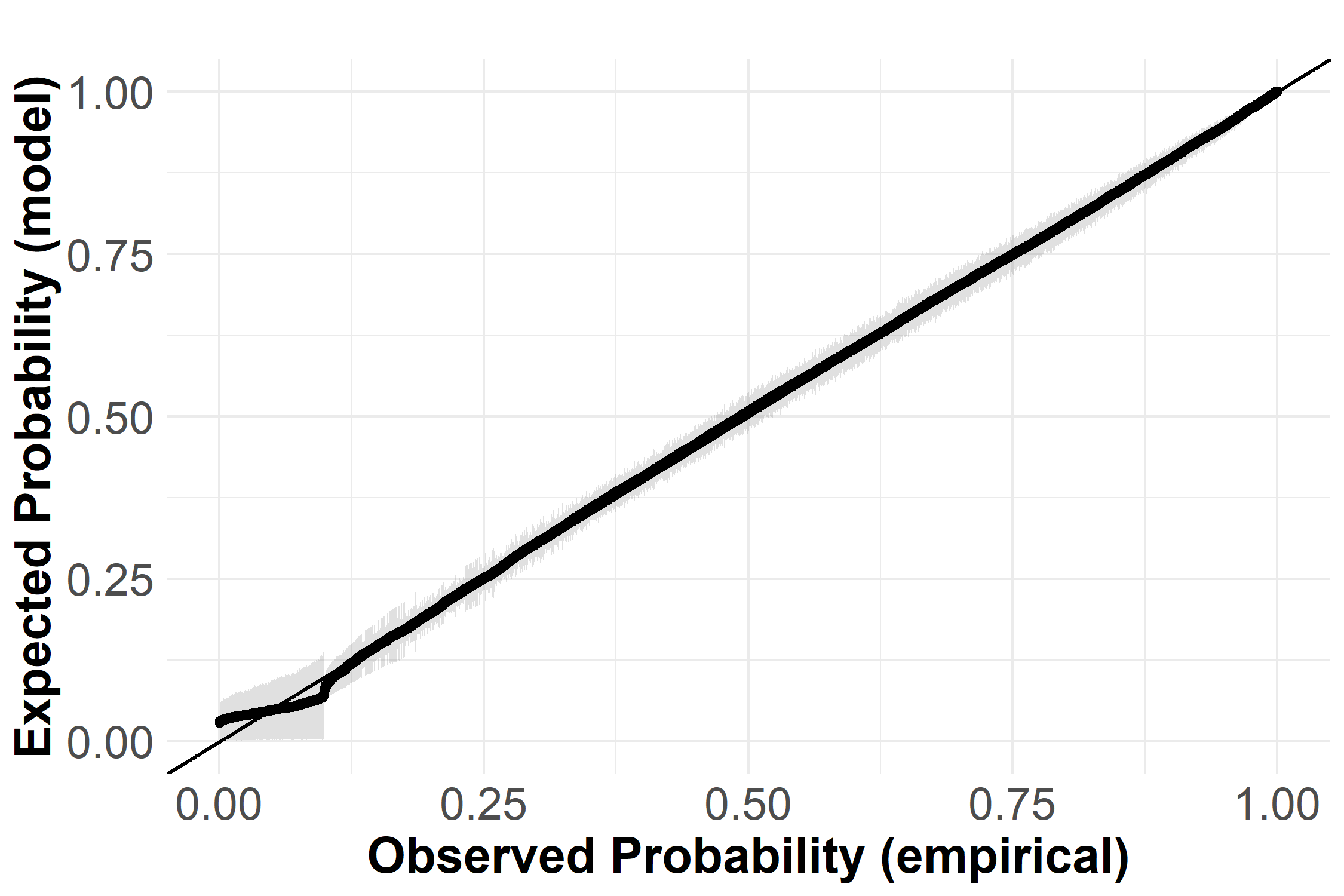}
    \includegraphics[width=0.32\linewidth]{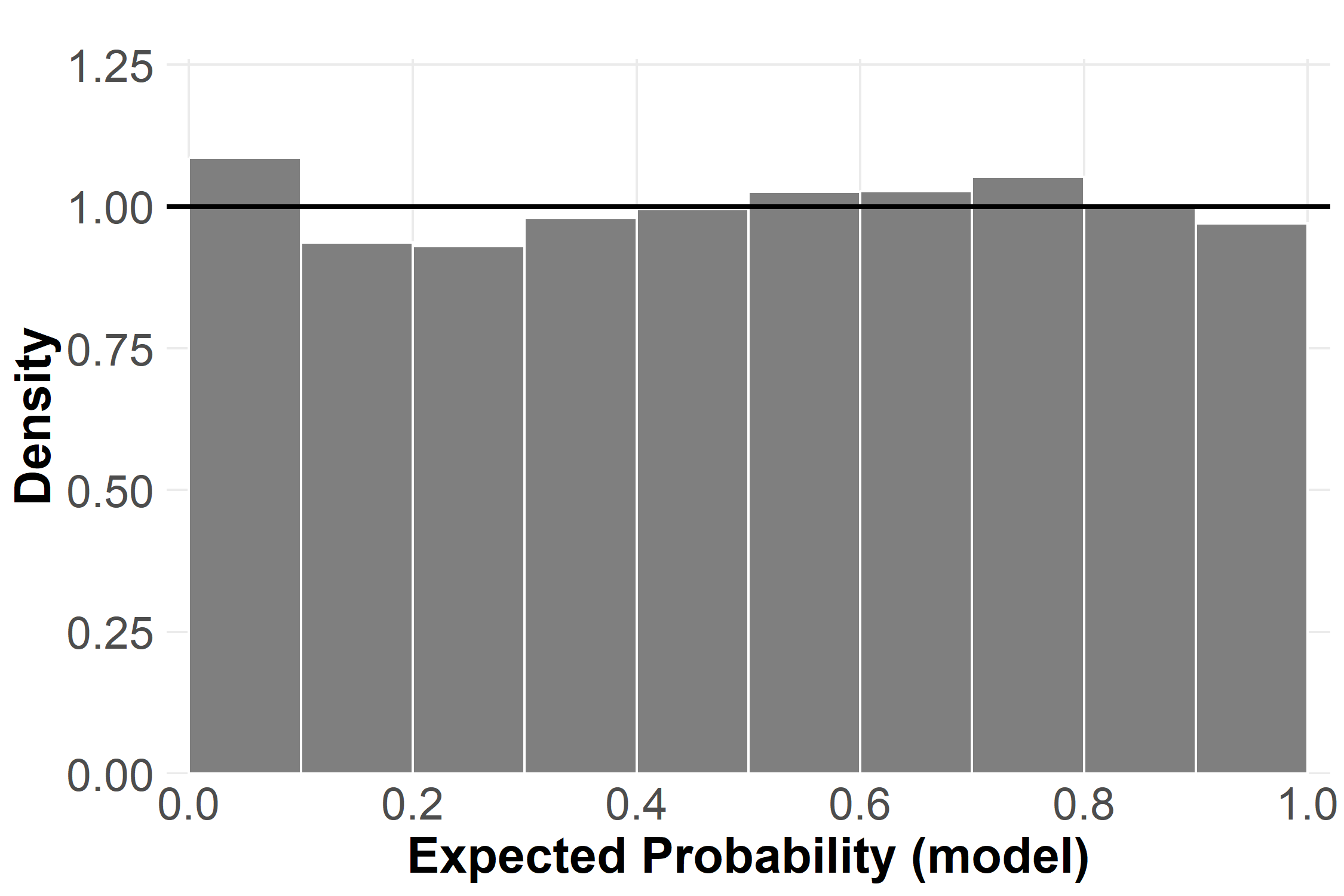}
    \includegraphics[width = 0.32\linewidth]{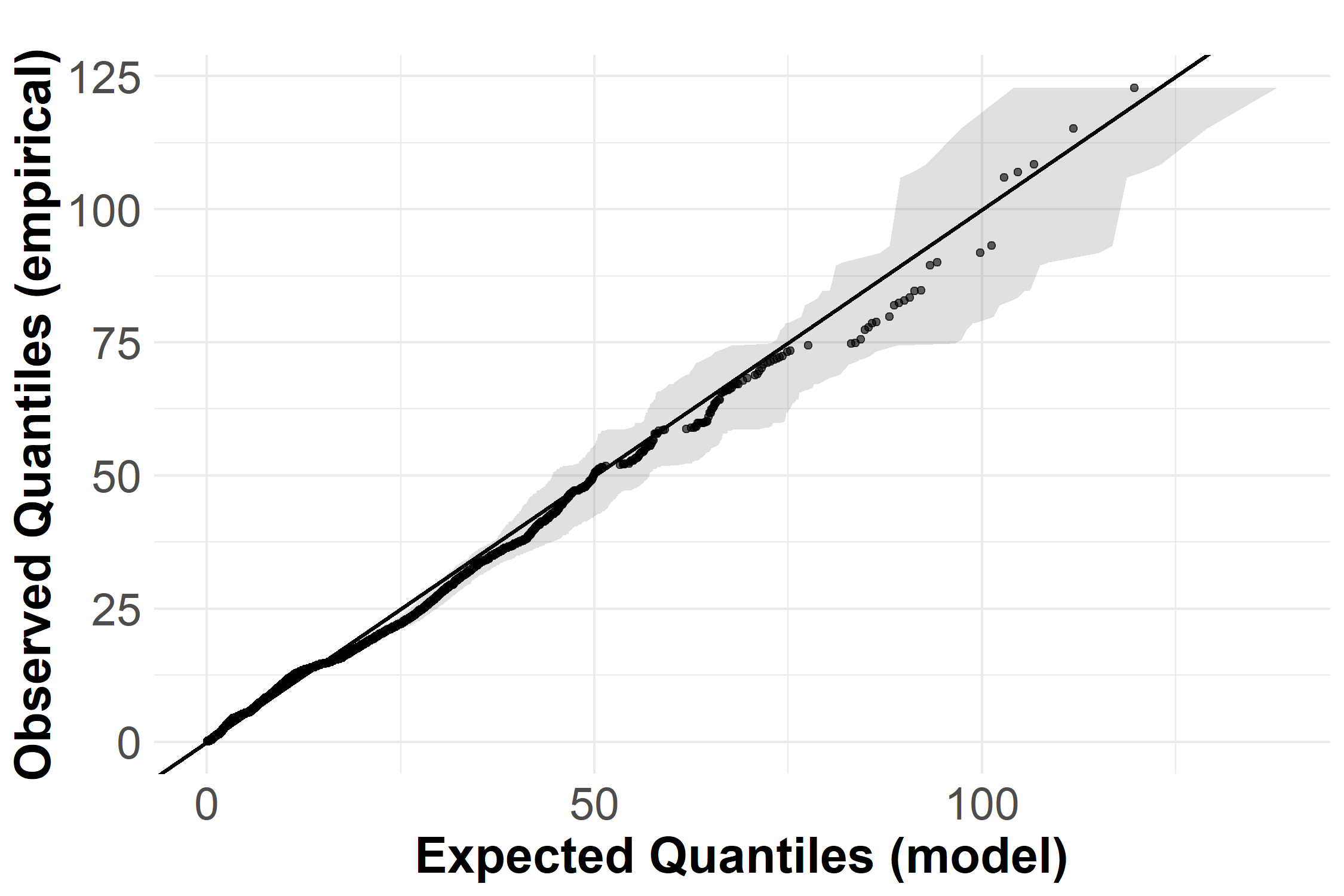}
    \caption{Model checking metrics for the chosen model for positive observations across the entire dataset. Left: PP-plot with 95\% credible interval. Center: PIT histogram. Right: QQ-plot with 95\% credible interval.}
    \label{pplot_PIT}
\end{figure}

After choosing this model, we compute the model checking tools introduced in Section~\ref{model.comp} to assess its calibration and sharpness with respect to the positive values available. Figure~\ref{pplot_PIT} shows the PP-plot, PIT histogram, and QQ-plot for the whole time period in all stations altogether. Results show very good agreement between the empirical and model probabilities, as well as for the quantiles, highlighting the model adequacy for the whole dataset. In Section~S4.2, QQ-plots conditional on the previous day precipitation indicator can be found, as well as separate plots for each station. 

\subsection{Results for the selected model}

For the selected model from the previous section, DZI-SCGLM,
we present a summary of the posterior distribution of the regression coefficients and parameters, and the results obtained after the spatial interpolation and sampling described in Section~\ref{inference_met}. To present the results, we use a fine grid $G_D$ of points within the region of interest $D$. The grid contains $\lvert G_D \rvert = 1320$ grid cells, which correspond to a resolution of $10$ km $\times$ $10$ km. Altogether, the entire posterior predictive dataset has $15$ years $\times$ $92$ days $\times$ $1320$ grid centroids $\times$ $1000$ replicates $\approx$ $1.8$ billion points.

\subsubsection{Model parameters}

Table~\ref{model_parameters} summarizes the posterior distribution of the regression coefficients for both the occurrence and intensity models. In both models, the $95\%$ credible intervals (CIs) for the effects of the NAO, the previous day precipitation indicator, and station elevation exclude zero and share the same sign across models. The intercept in the intensity model also excludes zero but displays the opposite sign compared to the occurrence model. The posterior distributions for both covariates related to the distance to the coast overlapped with zero. These findings deviate slightly from the initial exploratory analysis, where all covariates showed significant effects. The presence of non-credible terms in the final spatio-temporal DZI model can be attributed to spatial confounding, where the inclusion of spatial random effects influences and partially absorbs the signal of the spatial covariates, leading to wider CI that encompass zero \citep{paciorek2010}.

The dispersion parameter $\phi$ has a posterior mean and $95\%$ CI of $2.96$ $(2.89, \, 3.04)$. The precision parameters for the occurrence and intensity models exhibit a posterior mean and $95\%$ CI of $16.03$ $(10.66, \, 23.03)$ and $10.31$~$(6.57, \, 15.24)$, respectively. Finally, the decay parameters impose an effective range $3/\phi_\beta$ and $3/\phi_\gamma$ with mean and $95$\% CI of approximately $371$ $(246, \, 541)$ km and $308$ $(206, \, 447)$ km, respectively. 

\begin{table}[]
    \centering
    \begin{tabular}{l|cc|cc}
    \hline
     \multicolumn{1}{c|}{\multirow{2}{*}{\textbf{Coefficient of }}} & \multicolumn{2}{c|}{\textbf{Occurrence}} & \multicolumn{2}{c}{\textbf{Intensity}} \\
     \cmidrule(lr){2-3} \cmidrule(lr){4-5}
     \multicolumn{1}{c|}{} & Mean & $95$\% CI & Mean & $95$\% CI\\
    \hline 
    Intercept & $-0.35$ & $(-0.81, \, 0.15)$ & $\mathbf{1.27}$ & $(0.55, \, 1.96)$\\
    NAO$_t$ & $\mathbf{-0.11}$ & $(-0.13, \, -0.09)$ & $\mathbf{-0.09}$ & $(-0.12, \, -0.07)$\\
    $I_{t\ell}(\bs)$ & $\mathbf{1.15}$ & $(1.12, \, 1.19)$ & $\mathbf{0.40}$ & $(0.36, \, 0.44)$\\
    elevation$(\bs)$ (km) & $\mathbf{0.21}$ & $(0.06, \, 0.35)$ & $\mathbf{0.35}$ & $(0.16, \, 0.56)$\\
    $\log(1 + \text{dist Med}(\bs))$ & $0.01$ & $(-0.06, \, 0.07)$ & $-0.01$ & $(-0.11, \, 0.08)$\\
    $\log(1 + \text{dist Can}(\bs))$ & $-0.03$ & $(-0.09, \, 0.03)$ & $-0.05$ & $(-0.14, \, 0.04)$\\
    \hline
    \end{tabular}
    \caption{Posterior mean and $95$\% CI of the regression coefficients of the occurrence and intensity models for the selected DZI-SCGLM model. The mean of the coefficients whose CI does not contain zero is marked in bold.}
    \label{model_parameters}
\end{table}

\begin{figure}[t]
    \centering
    \includegraphics[width=0.47\linewidth]{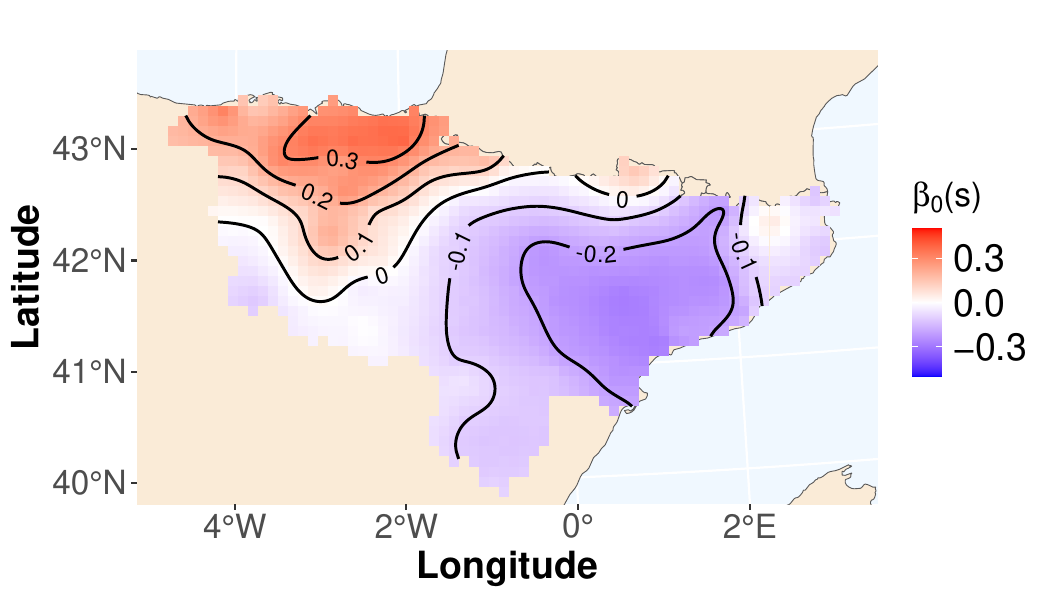}
    \includegraphics[width = 0.47\textwidth]{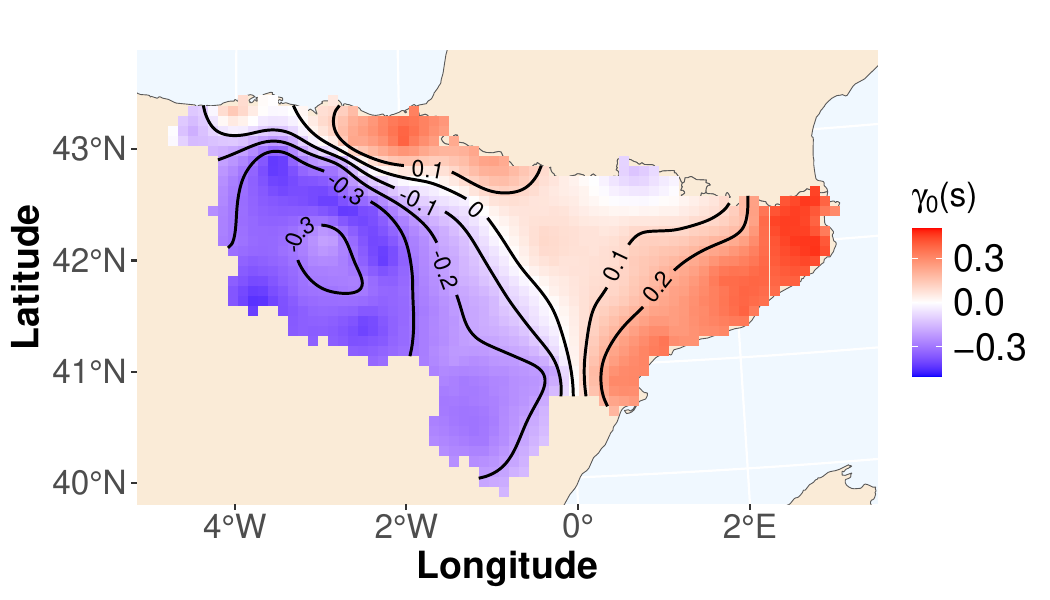}
    \caption{Left: Random mean surface of $\beta_0(\bs)$ (occurrence model). Right: Random mean surface of $\gamma_0(\bs)$ (intensity model).}
    \label{REsurf}
\end{figure}

Next, we look at the random effects posterior mean surface for the occurrence and intensity models (see Figure~\ref{REsurf}), calculated via Bayesian kriging. These effects provide local adjustment to the global intercept and spatial covariates.
For the occurrence model (left plot), where the probability of a wet event is modeled, we can see that the northwestern part of the region exhibits a positive random effect, resulting in an increase of the probability of a wet event in the whole study period. On the contrary, the eastern part of the region experiences a decrease in the probability of a wet event. In the case of a wet event, the right plot shows the spatial effect in the intensity model. As we can see, the Mediterranean coast, and part of the Pyrenees and Cantabrian coast, have a positive effect, resulting in an increase in precipitation intensity when it is positive. In the western and central part of the region, the spatial effect is mainly negative, indicating a decreasing effect in the intensity of precipitation. 

Furthermore, sensitivity analyses performed across a variety of sensible prior distributions demonstrated considerable robustness. The only exception occurred within the covariance hyperparameters of the Gaussian processes, which exhibited sensitivity to prior choices, consistent with the well-known lack of identifiability of spatial range and variance parameters \citep[][Chapter~6]{zhang2004,banerjee2025}. Crucially, overall model inference remained unchanged, and the CV metrics were equivalent regardless of the prior specification. As could be expected, adopting a much shorter effective spatial range made the distance to the coast covariates significant; restricting the flexibility of the random effects prevented them from absorbing the covariate signals. Based on these insights, we retained a final configuration with a physically reasonable effective spatial range scaled to the dimensions of our study region.

\subsubsection{Inference results}
Now, we show the results for the model-based inference metrics. Those shown here consist of some of the marginal metrics averaged across the whole study period (MAM 2010--2024). Due to the transition of detection levels from $0.1$ to $0.2$ in the measurement devices (see Figure~\ref{fig:map}), we show the results of considering a fixed detection level of $\epsilon = 0.2$ across space and time. Results for the marginal metrics when $\epsilon = 0.1$ may be found in Figure~S33. While the spatial behavior remains unaffected, lowering the detection limit reduces the effects on undetected precipitation and censoring, which is consistent with expectations. Refer to Table \ref{table_metrics} for the notation and brief description of the metrics.

\begin{figure}[t]
    \centering
    \includegraphics[width=0.47\linewidth]{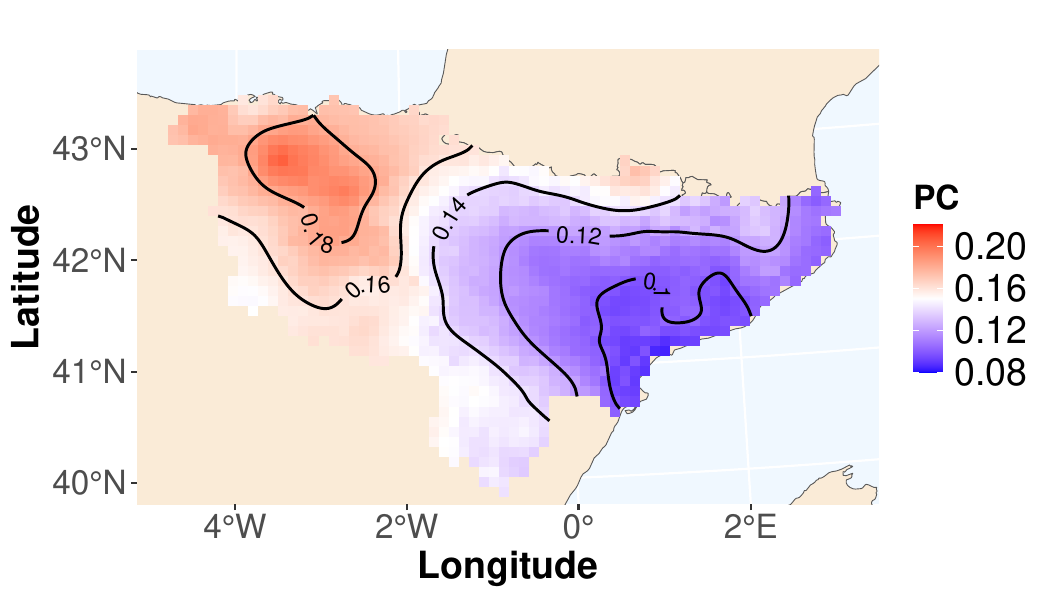}
    \includegraphics[width=0.47\linewidth]{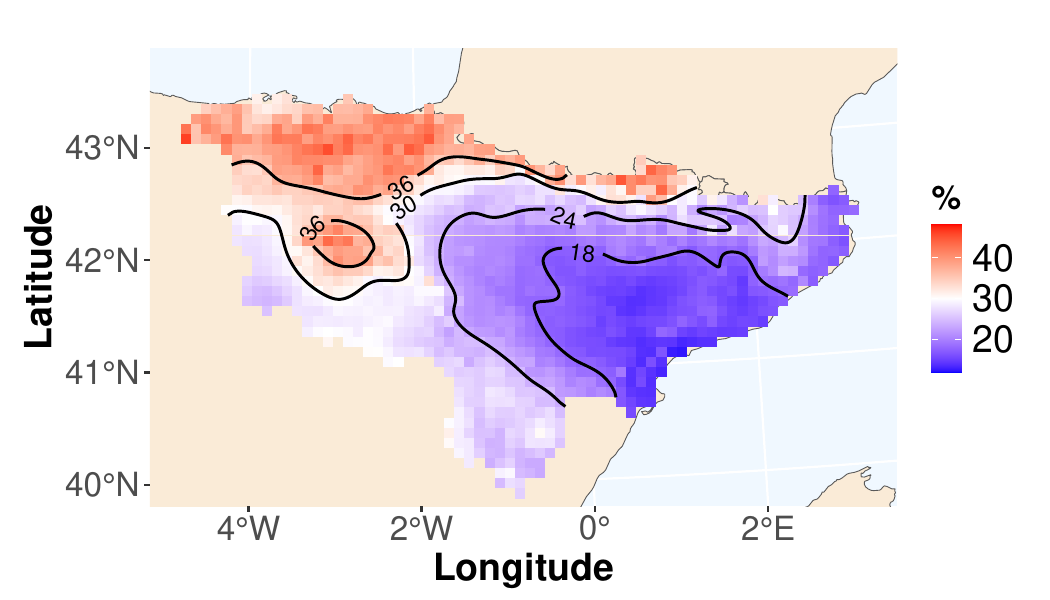}
    \caption{Left: Average probability of censoring (PC) in the whole study period when the detection level is set to $\epsilon = 0.2$ and conditions for precipitation exist. Right: Average percentage of censored days out of days with observed zeros (PCD) in the whole study period when the detection level is set to $\epsilon = 0.2$.}
    \label{prob_cens_0.2}
\end{figure}

\paragraph{Probability of censored precipitation.}
Figure~\ref{prob_cens_0.2} shows the average PC (left plot) and PCD (right plot) values for the MAM 2010--2024 period. In the eastern part of the study region, the PC is lower, decreasing from 0.14 to 0.08 toward the Mediterranean coast. In contrast, the western part exhibits higher and more spatially homogeneous PC, ranging from 0.16 to 0.18 toward the northern areas. This indicates that the northwestern sector, which is characterized by a temperate Atlantic climate with frequent precipitation and high humidity throughout much of the year, experiences a higher frequency of wet events with precipitation below the detection limit compared to the eastern sector of the Ebro River Basin, where such events are less frequent.

The computation of the average PCD directly addresses a key objective of the proposed methodology: quantifying the effect of the detection threshold on the measurement devices. The northwestern part is the most affected, having between $30$\% and $45$\% of censored days out of days that with an observed zero. As noted above, the PC is higher in this part of the region. This implies that, to obtain a high proportion of censored values, the probability of observing zeros must be relatively low. The results show that observed zeros are unlikely to correspond to true dry days. This pattern suggests a wetter precipitation regime in this area, characterized by frequent low-precipitation events, consistent with the climatic conditions of the region. In contrast, the central and eastern parts of the region appear to be less affected by censoring, with estimated PCD ranging from 15\% to 24\%. This corresponds to a higher probability that observed zeros represent actual dry days, matching the drier Mediterranean and continental-influenced climate regimes of the area, where drizzle events are rare.

It is also important to mention that the mean range of the $95$\% CI is approximately $17\%$, being consistent in most of the study region. The highest range values are found in the underrepresented part of the Pyrenees and the bottom border of the defined grid; see Figure S30.

\begin{figure}[t]
    \centering
    \includegraphics[width=0.47\linewidth]{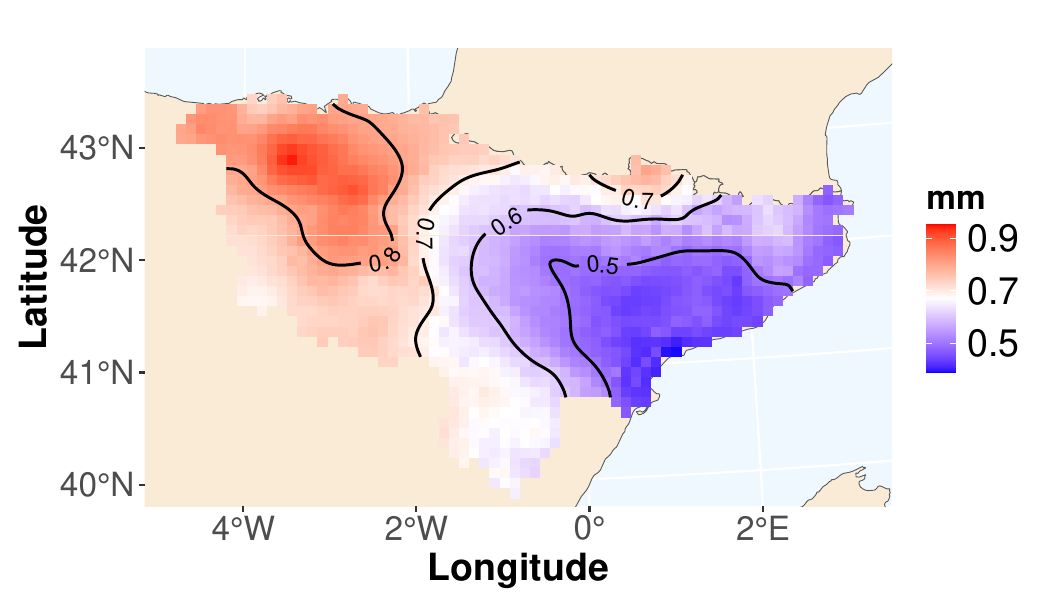}
    \includegraphics[width = 0.47\linewidth]{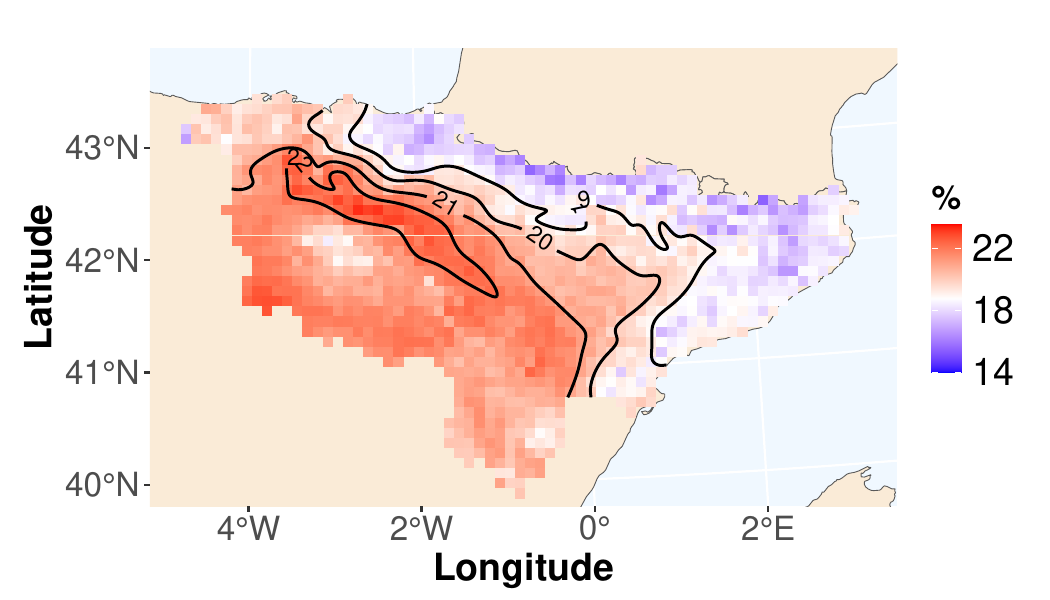}
    \caption{Left: Average yearly March to May expected undetected precipitation (EUP) in the whole study period when the detection level is set to $\epsilon = 0.2$. Right: Percentage difference of true precipitation relative to observed precipitation of the $0.90$ quantile (PRDQ90) in the whole study period when the detection level is set to $\epsilon = 0.2$.}
    \label{perc_and_HR}
\end{figure}

\paragraph{Expectation of undetected precipitation.} Based on the preceding findings, the next step is to quantify the actual volume of precipitation that has been undetected. The left plot in Figure~\ref{perc_and_HR} shows an accumulated EUP value in MAM months for each year, which gives us insight into how much precipitation has gone undetected on average each MAM of the 15-year period. Although the EUP values do not seem high, they highlight the effect of detection limits. As before, the northwestern part of the region exhibits the highest values, with averages between $0.7$ and $0.9$ mm of yearly MAM undetected precipitation. On the contrary, the Mediterranean coast has a lesser amount of yearly MAM EUP, with values less than $0.5$ mm. The mean range of the $95$\% CI is approximately $0.3$; see Figure~S31. 

Although detection limits may not have a substantial effect on total precipitation estimates, they may introduce biases in the estimation of quantiles, specially the extreme ones. As an illustration, the right plot of Figure~\ref{perc_and_HR} shows the average PRDQ90 values in the 15-year period. The results show how the $0.90$ quantile of positive precipitation is clearly affected by the detection limit. 
In the Pyrenees and Mediterranean coast, the $0.90$ observed quantile overestimated the true one, being the true quantile approximately $19$\% lower than the observed. On the other hand, this overestimation is slightly higher in the central part of the Ebro River Basin, with true quantiles being around $21$\% lower than the observed. In the case of studying this difference in the mean, there exists a negligible difference; see Figure S32. 

\paragraph{Metrics conditional on previous day precipitation indicator.} The previous results correspond to the marginal metrics. As mentioned at the end of Section~\ref{inference_met}, these metrics may be computed conditionally on the previous day precipitation indicator.

\begin{figure}
    \centering
    \includegraphics[width=0.47\linewidth]{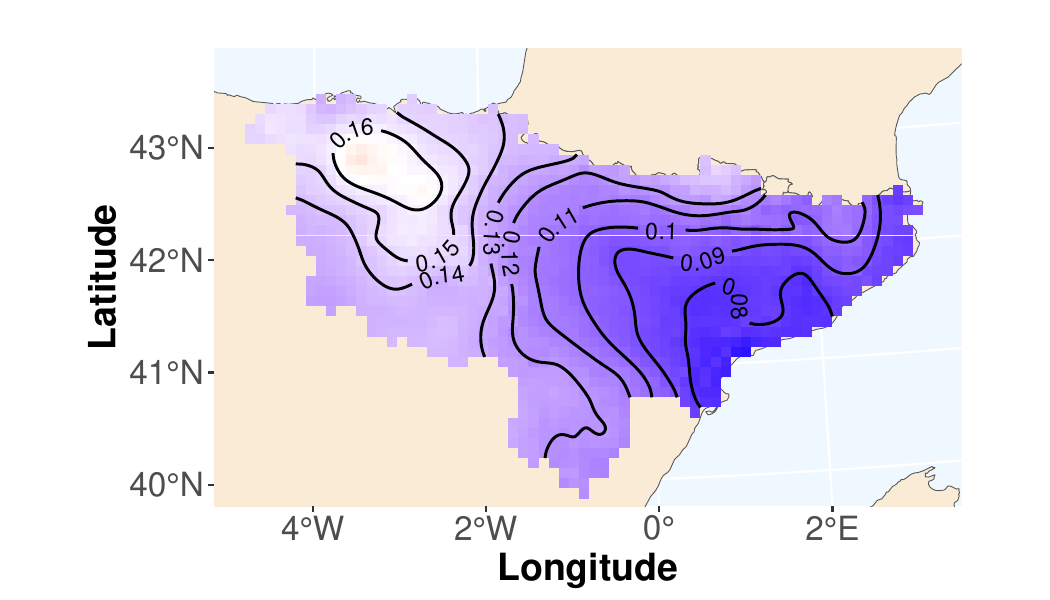}
    \includegraphics[width=0.47\linewidth]{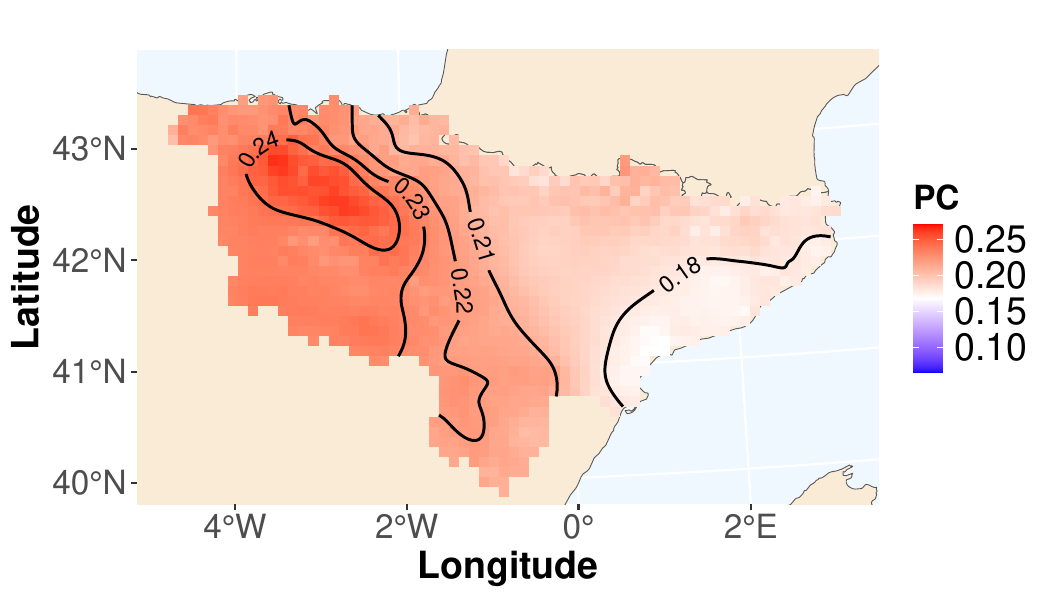}
    \caption{Average probability of censoring (PC) in the whole study period conditional on the previous day precipitation indicator being equal to 0 (left) or 1 (right) when the detection level is set to $\epsilon = 0.2$.}
    \label{prob_cens_cond}
\end{figure}

Figure \ref{prob_cens_cond} shows the average conditional PC. The conditional probabilities of censoring exhibit different spatial behaviors and magnitudes when the previous day precipitation indicator is equal to $0$ (left) or $1$ (right). When the previous day has not exceeded $1$ mm, the PC decreases from $0.16$ to $0.07$ as we move from the Cantabrian coast to the Mediterranean coast. On the other hand, when the previous day exceeded $1$ mm, the magnitude of the PC increases considerably, and the spatial distribution slightly changes. The northwestern part of the region also exhibits the greatest values, with an increase of approximately $0.08$ with respect to the values conditional on the previous day not exceeding $1$ mm. This increase is higher in the Mediterranean coast, where we find increases up to $0.09$. Additional results regarding the remaining metrics may be found in Figure S34, as well as the results when the detection limit is set to $\epsilon = 0.1$ in Figure S35.

\section{Summary and future work }\label{conclusion}

We have taken up the novel challenge of explaining the incidence of two types of zeros arising in data with detection limits.  One type of zero arises through unsuitability for a zero at the day, year and location of an observation.  The other zero arises through the minimum detection level of the measuring device employed in that year, on that day, at that location. We have specified flexible DZI models to explain both types of zeros as well as the detection-level censored positive observations. By incorporating both fixed and random effects, these models successfully capture the spatial patterns in the probability of censoring and undetected precipitation, allowing us to explicitly identify and separate both sources of zeros. Specifically, we have supplied a flexible multi-level model and fitted it within a Bayesian framework to offer rich inference  with exact uncertainty under the model.  We have applied the modeling approach to a dataset consisting of daily precipitation levels for 15 years at 70 monitoring stations in the northeastern part of Spain.  We have examined covariate selection, model adequacy and comparison, and reported numerous interesting findings.

The results for the study area around the Ebro River Basin highlight the importance of adequately accounting for the probability of censoring, and the associated undetected precipitation due to the detection limits of measurement devices. For a detection threshold of $\epsilon=0.2$, which is currently used in many Spanish gauging stations, the percentage of ``false'' zeros ranges from 30\% to 45\% of observed zeros in the northwestern part of the region, making it the most affected area. In the central and eastern areas, this proportion ranges from 15\% to 24\%. 
Although detection limits may not substantially affect estimates of total precipitation, they introduce biases in the estimation of precipitation quantiles, particularly in the upper tail of the distribution. We find that the observed $0.90$ quantile overestimates the true quantile, with the true quantile being approximately 19\% to 21\% lower than the observed value across the study region.

In summary, the proposed model enables the assessment of biases affecting several precipitation characteristics, including the number of wet days, the length of wet spells, drought indices, and precipitation quantiles. The achieved estimation of light precipitation may be of relevance for sensitive ecosystems and evapotranspiration and infiltration models. In addition, the model facilitates the consistent combination of stations and time periods with different detection limits, thereby avoiding biases in spatial and temporal analyses.

It is worth noting that in adopting a Gamma distribution for positive precipitation, we have preferred this distributional shape to say a log normal or a tobit normal.  Neither of these latter two models would provide tail behavior we believe to be more appropriate.  However, a heavier tailed model for the positive precipitation, e.g., a model which incorporates a Gamma distribution for the bulk of the positive data along with an extreme value distribution or an extended generalized Pareto distribution to better capture the right tail is a possibility for future work. In this regard, introducing a spatially varying dispersion parameter for the Gamma distribution, $\phi(\bs)$, could be of interest, though quite ambitious.  Other future work includes the development of quantile models for precipitation, again with a point mass at zero.  Also of interest might be investigation of precipitation amounts at different temporal resolutions where there are still many zeros, as well as considering more than just previous day dependence in specifying the distribution for the amount of today's precipitation.

\section*{Acknowledgments}
This work was partially supported by MCIN/AEI/10.13039/501100011033 and Unión Europea NextGenerationEU under Grant PID2023-150234NB-I00, and BIOSTATNET network RED2024-153680-T; and Gobierno de Aragón under Grants PROY\_T21\_24, UNIZAR 15075, and Research Group `Modelos Estocásticos'. J. M.-G. was supported by Gobierno de Aragón under Doctoral Scholarship ORDEN EMC/590/2025.

\appendix

\section{Metropolis-within-Gibbs sampler} \label{MCMC}
This appendix develops the MCMC model fitting algorithm in detail; in particular, it describes the full conditional distributions needed for block Gibbs sampling and guidelines for sampling from each block of parameters. First, it is convenient to denote parameters in vector notation as follows. Let the vector of observed precipitations be:
\begin{equation*}
    \mathbf{y}^\text{obs} = ({\mathbf{y}_{1,1}^\text{obs}}^{\top},\ldots,{\mathbf{y}_{TL}^\text{obs}}^{\top})^{\top} \quad \text{with} \quad \mathbf{y}_{t\ell}^\text{obs} = (y_{t\ell}^\text{obs}(\bs_1),\ldots,y_{t\ell}^\text{obs}(\bs_n))^{\top};
\end{equation*}
equivalently, define the vectors $\mathbf{Y}^\text{true}$, $\mathbf{Y}_{t\ell}^\text{true}$, $\mathbf{Z}$, and $\mathbf{Z}_{t\ell}$ from $Y_{t\ell}^{\text{true}}(\bs_i)$ and $Z_{t\ell}(\bs_i)$, respectively. Implementing hierarchical centering of the global intercept and spatial covariates over the spatial Gaussian processes \citep{gelfand1995}, and overloading the notation introduced so far, we have:
\begin{align*}
    \bm{\beta}_0 &= (\beta_0(\mathbf{s}_1), \dots, \beta_0(\mathbf{s}_n))^\top \sim N_n\left(\tilde{\mathbf{X}}  \tilde{\bm\beta}, \sigma^2_{\beta} \mathbf{R}_{\phi_\beta} \right),
    \\ \quad \bm{\gamma}_0 &= (\gamma_0(\mathbf{s}_1), \dots, \gamma_0(\mathbf{s}_n))^\top \sim N_n\left(\tilde{\mathbf{X}} \tilde{\bm\gamma}, \sigma^2_{\gamma} \mathbf{R}_{\phi_\gamma} \right),
\end{align*}
where $\tilde{\mathbf{X}} = (\tilde{\mathbf{x}} (\mathbf{s}_1), \dots, \tilde{\mathbf{x}} (\mathbf{s}_n))^{\top}$ corresponds to the $n \times k$ design matrix containing the global intercept and the $k - 1$ spatial variables, $\tilde{\bm\beta}$ and $\tilde{\bm\gamma}$ denote the $k$-dimensional vectors of regression coefficients of the mean, and  $\mathbf{R}_{\phi_\beta}$ and $\mathbf{R}_{\phi_\gamma}$ are the $n \times n$ correlation matrices dependent on their respective decay parameter. To avoid introducing excessive notation, the $n$-dimensional spatial vectors $\bm{\beta}_0$ and $\bm{\gamma}_0$ are appropriately expanded to $TLn$-dimensional vectors where necessary. For the remainder set of covariates, we define the $(TLn) \times (p-k)$ design matrix $\mathbf{X} =(\mathbf{X}_{1,1},\ldots,\mathbf{X}_{TL})^{\top}$ with $\mathbf{X}_{t\ell} = (\mathbf{x}_{t\ell}(\bs_1),\ldots,\mathbf{x}_{t\ell}(\bs_n))^{\top}$. Analogously, let $\mathbf{Y}^{\text{true}}_+$ denote the subvector of positive true responses, i.e., $Y_{t\ell}^{\text{true}}(\bs) \mid Y_{t\ell}^{\text{true}}(\bs) > 0$, and $\mathbf{X}_{+}$ the corresponding restricted design matrix. Again, let $\bm{\theta}$ denote the spatial processes, model parameters, and hyperparameters. Let $\mathbf{1}_n$ denote the $n$-dimensional vector of ones. Finally, $G(\,\cdot \mid \mu, \phi)$ and $F_G(\,\cdot \mid \mu, \phi)$ denote the pdf and cdf, respectively, of a Gamma distribution with mean $\mu$ and dispersion $\phi$; however, unless otherwise specified, we employ the shape-rate parameterization for all Gamma prior and full conditional distributions.

    \paragraph{Joint full conditional distribution of $\mathbf{Z}$ and $\mathbf{Y}^\text{true}$.} We jointly sample the pair $(Z_{t\ell}(\bs_i), Y_{t\ell}^\text{true}(\bs_i))$ for every $t=1, \dots, T$, $\ell = 1, \dots, L$, and $i = 1, \dots, n$, from their joint full conditional distribution using:
    \begin{equation*}
    [\mathbf{Z}, \mathbf{Y}^{\text{true}} \mid \mathbf{y}^{\text{obs}}, \bm{\theta}] = [\mathbf{Y}^{\text{true}} \mid \mathbf{y}^{\text{obs}}, \bm{\theta}] \, [\mathbf{Z} \mid \mathbf{Y}^{\text{true}}, \bm{\theta}].
    \end{equation*}
    There are three possibilities. When the observed value is not zero, the true value is equal to the observed.
    
    When  $y_{t\ell}^\text{obs}(\bs_i) = 0$, a value of $Y_{t\ell}^\text{true}(\bs_i)$ is sampled from
    {\small
    \begin{align*}
    [Y_{t\ell}^{\text{true}}(\mathbf{s}_i) \mid y_{t\ell}^{\text{obs}}(\mathbf{s}_i) = 0, \bm{\theta}] &= 
    w_{t\ell} (\mathbf{s}_i) \cdot \delta_0(Y_{t\ell}^{\text{true}}(\mathbf{s}_i)) \\ 
    & \hspace{5mm} + (1 - w_{t\ell} (\mathbf{s}_i)) \cdot 
    \frac{G(Y_{t\ell}^{\text{true}}(\mathbf{s}_i) \mid \mu_{t\ell}(\mathbf{s}_i), \phi)}{\int_{0}^{\epsilon_{t\ell}(\mathbf{s}_i)} G(y \mid \mu_{t\ell}(\mathbf{s}_i), \phi) \, dy} \, \mathbf{1}\{0 < Y_{t\ell}^{\text{true}}(\mathbf{s}_i) < \epsilon_{t\ell}(\mathbf{s}_i)\},
    \end{align*}
    }
    with weight
    \begin{equation*}
        w_{t\ell}(\mathbf{s}_i) = \frac{1 - \Phi(\eta_{{t\ell}}(\mathbf{s}_i))}{1 - \Phi(\eta_{{t\ell}}(\mathbf{s}_i)) + \Phi(\eta_{{t\ell}}(\mathbf{s}_i)) \cdot \int_{0}^{\epsilon_{t\ell}(\mathbf{s}_i)} G(y \mid \mu_{{t\ell}}(\mathbf{s}_i), \phi) \, dy}.
    \end{equation*} 
    
     When $y_{t\ell}^\text{obs}(\bs_i)$ is missing and a previous day precipitation indicator of exceeding $1$ mm, $I_{t\ell}(\bs_i)$, has been included in the model setting, a value of $Y_{t\ell}^\text{true}(\bs_i)$ is sampled from 
    \begin{align*}
    [Y_{t\ell}^{\text{true}}(\bs_i) \mid \bm\theta] &\propto [Y_{t\ell}^{\text{true}}(\bs_i) \mid Y_{t,\ell-1}^{\text{true}}(\bs_i), \bm\theta][Y_{t,\ell+1}^{\text{true}}(\bs_i)\mid Y_{t\ell}^{\text{true}}(\bs_i), \bm\theta] \\
    &\propto P^0_{t\ell}(\bs_i) \cdot \delta_0(Y_{t\ell}^{\text{true}}(\bs_i)) + P^{(0, 1]}_{t\ell}(\bs_i) \cdot \frac{G(Y_{t\ell}^{\text{true}}(\bs_i) \mid \mu_{t\ell}(\bs_i), \phi)}{F_G(1 \mid \mu_{t\ell}(\bs_i), \phi)} \mathbf{1}{\{0 < Y_{t\ell}^{\text{true}}(\bs_i) \leq 1\}} \\
    & \hspace{.5cm}+ P^{(1, \infty)}_{t\ell}(\bs_i) \cdot \frac{G(Y_{t\ell}^{\text{true}}(\bs_i) \mid \mu_{t\ell}(\bs_i), \phi)}{1 -  F_G(1 \mid \mu_{t\ell}(\bs_i), \phi)} \mathbf{1}{\{Y_{t\ell}^{\text{true}}(\bs_i) > 1\}} , 
    \end{align*}
    with 
    \begin{align*}
        P^0_{t\ell}(\bs_i) = \frac{w^0_{t\ell}(\bs_i)}{W_{t\ell}(\bs_i)}, \quad P^{(0, 1]}_{t\ell}(\bs_i) = \frac{w^{(0, 1]}_{t\ell}(\bs_i)}{W_{t\ell}(\bs_i)}, \quad P^{(1, \infty)}_{t\ell}(\bs_i) = \frac{w^{(1, \infty)}_{t\ell}(\bs_i)}{W_{t\ell}(\bs_i)},
    \end{align*}
    where $W_{t\ell}(\bs_i) = w^0_{t\ell}(\bs_i) + w^{(0, 1]}_{t\ell}(\bs_i) + w^{(1, \infty)}_{t\ell}(\bs_i)$ with weights
    {
    \scriptsize
    \begin{align*}
        w_0  &= \begin{cases}
            \pi_{t\ell}(\bs_i)  \cdot \pi_{t,\ell + 1}^{(0)}(\bs_i), & \text{if} \quad Y_{t,\ell + 1}^{\text{true}}(\bs_i) = 0, \\
            \pi_{t\ell}(\bs_i) \cdot (1 - \pi_{t,\ell+1}^{(0)}(\bs_i))\cdot G(Y_{t,\ell+1}^{\text{true}}(\bs_i) \mid \mu_{t,\ell + 1}^{(0)}(\bs_i), \phi ), & \text{if} \quad Y_{t,\ell + 1}^{\text{true}}(\bs_i) > 0, \\
        \end{cases} \\
        w_{(0, 1]} & = \begin{cases}
            (1 - \pi_{t\ell}(\bs_i)) \cdot F_G(1 \mid \mu_{t\ell}(\bs_i), \phi) \cdot \pi_{t,\ell+1}^{(0)}(\bs_i), & \text{if} \quad Y_{t,\ell + 1}^{\text{true}}(\bs_i) = 0, \\
            (1 - \pi_{t\ell}(\bs_i)) \cdot F_G(1 \mid \mu_{t\ell}(\bs_i), \phi) \cdot (1 - \pi_{t,\ell+1}^{(0)}(\bs_i)) \cdot G(Y_{t,\ell+1}^{\text{true}}(\bs_i) \mid \mu_{t,\ell+1}^{(0)}(\bs_i), \phi), & \text{if} \quad Y_{t,\ell + 1}^{\text{true}}(\bs_i) > 0, 
        \end{cases} \\
        w_{(1, \infty)} & = \begin{cases}
            (1 - \pi_{t\ell}(\bs_i)) \cdot (1 - F_G(1 \mid \mu_{t\ell}(\bs_i), \phi)) \cdot \pi_{t,\ell+1}^{(1)}(\bs_i), & \text{if} \quad Y_{t,\ell + 1}^{\text{true}}(\bs_i) = 0, \\
            (1 - \pi_{t\ell}(\bs_i)) \cdot (1 - F_G(1 \mid \mu_{t\ell}(\bs_i), \phi)) \cdot (1 - \pi_{t,\ell+1}^{(1)}(\bs_i)) \cdot G(Y_{t,\ell+1}^{\text{true}}(\bs_i) \mid \mu_{t,\ell+1}^{(1)}(\bs_i), \phi), & \text{if} \quad Y_{t,\ell + 1}^{\text{true}}(\bs_i) > 0, 
        \end{cases} 
    \end{align*}}
    where the superscripts $(1)$ and $(0)$ indicate whether or not the estimated coefficient for the previous day's indicator, $I_{t\ell}(\bs_i)$, is included. The sampling of values from a truncated  Gamma distribution was performed following the method proposed by \cite{philippe1997}.
    
    After having sampled a value of $Y_{t\ell}^\text{true}(\bs_i)$, we sample $Z_{t\ell}(\bs_i)$ from its truncated normal following the method proposed by \cite{robert1995}, 
    \begin{equation*}
    [Z_{t\ell}(\mathbf{s}_i) \mid Y_{t\ell}^{\text{true}}(\mathbf{s}_i), \bm{\theta}] \propto 
    \begin{cases}
    {N}(Z_{t\ell}(\mathbf{s}_i) \mid \eta_{{t\ell}}(\mathbf{s}_i), 1) \cdot \mathbf{1}{\{Z_{t\ell}(\mathbf{s}_i) \le 0\}}, & \text{if } Y^{\text{true}}_{t\ell}(\mathbf{s}_i) = 0, \\
    {N}(Z_{t\ell}(\mathbf{s}_i) \mid \eta_{{t\ell}}(\mathbf{s}_i), 1) \cdot \mathbf{1}{\{Z_{t\ell}(\mathbf{s}_i) > 0\}}, & \text{if } Y^{\text{true}}_{t\ell}(\mathbf{s}_i) > 0.
    \end{cases}
    \end{equation*}
    
    \paragraph{Full conditional distribution of $\bm\beta_0$ and $\bm\beta$.}
    We sample $\bm\beta_0$ from its normal full conditional distribution,
    \begin{align*}
        [\bm{\beta}_0 \mid \bm\beta, \mathbf{Z}] &= N_n(\bm\beta_0 \mid \mathbf{m}_{{\beta}_0}, \mathbf{C}_{{\beta}_0}); \\
        \mathbf{C}_{{\beta}_0} & = \left( T L \mathbf{I}_n + \frac{1}{\sigma_{\beta} ^2} \mathbf{R}_{\phi_{\beta}}^{-1}\right) ^{-1},\\
        \mathbf{m}_{{\beta}_0} & =  \mathbf{C}_{{\beta}_0} \left(\sum_{t, \ell} (\mathbf{Z}_{t\ell} - \mathbf{X}_{t \ell}\bm\beta) 
        + \frac{1}{\sigma_{\beta}^2} \mathbf{R}_{\phi_{\beta}}^{-1} \tilde{\mathbf{X}} \tilde{\bm{\beta}} \right).
    \end{align*} 
    
    Given the normal prior for the regression coefficients,  $\bm{\beta} \sim N_{p-k}( \mathbf{a}_{{\beta}}, \mathbf{B}_{{\beta}})$, we sample them from their normal full conditional distribution,
\begin{align*}
    [\bm{\beta} \mid \bm\beta_0, \mathbf{Z}] &= N_{p-k}(\bm{\beta} \mid \mathbf{m}_{{\beta}} , \mathbf{C}_{{\beta}} ); \\
    \mathbf{C}_{{\beta}} &= \left( \mathbf{X}^\top \mathbf{X} + \mathbf{B}_{{\beta}}^{-1} \right)^{-1}, \\
    \mathbf{m}_{{\beta}} &= \mathbf{C}_{{\beta}} \left( \mathbf{X}^\top (\mathbf{Z} - \bm{\beta}_0) + \mathbf{B}_{{\beta}}^{-1} \mathbf{a}_{{\beta}} \right).
\end{align*}

    \paragraph{Full conditional distribution of $\bm\gamma_0$, $\bm\gamma$ and $\phi$.}
    Following \cite{gamerman1997}, we define the working response as 
    \begin{equation*}
    \tilde{\mathbf{Y}}= \mathbf{Y}^{\text{true}}_+ \oslash \exp(\mathbf{X}_+ \bm{\gamma} + \bm{\gamma}_0) +   \mathbf{X}_+ \bm{\gamma} + \bm{\gamma}_0 - \mathbf{1}_{N_+},
    \end{equation*}
     where $\oslash$ denotes element-wise (Hadamard) division and $N_{+} = \sum_{t, \ell, i} \mathbf{1}\{ Y_{t\ell}^\text{true}(\bs_i) > 0\}$ is the total number of positive true responses. 
    We sample values for $\bm\gamma_0$ from its non-standard full conditional posterior distribution
    \begin{align*}
        [\bm{\gamma}_0 \mid \mathbf{Y}^{\text{true}}_+, \bm\gamma, \phi] 
        & \propto
        \exp \left\{ -\frac{1}{2 \sigma_\gamma^2} \left( \bm{\gamma}_0 -  \tilde{\mathbf{X}} \tilde{\bm{\gamma}} \right)^\top \mathbf{R}^{-1}_{\phi_{\gamma}} \left( \bm{\gamma}_0 - \tilde{\mathbf{X}} \tilde{\bm{\gamma}}\right) -
        \frac{1}{\phi} \mathbf{1}^\top_{N_+} \left( \tilde{\mathbf{Y}} -  \mathbf{X}_+ \bm{\gamma} + \mathbf{1}_{N_+} \right) \right\}.
    \end{align*}
    We update the spatial effects values incorporating an iterative weighted least squares within a Metropolis-Hastings \citep{gamerman1997} step with the following multivariate normal proposal for a candidate $\bm{\gamma}_0^*$,
    \begin{align*}
        q(\bm{\gamma}^*_0 \mid \bm{\gamma}_0) &= N_n(\bm{\gamma}^*_0 \mid \mathbf{m}_{{\gamma}_0} , \mathbf{C}_{{\gamma}_0} ); \\
        \mathbf{C}_{{\gamma}_0} &= \left( \frac{1}{\phi} \bm\Omega + \frac{1}{\sigma_{\gamma}^2}\mathbf{R}^{-1}_{\phi_\gamma} \right)^{-1}, \\
        \mathbf{m}_{{\gamma}_0} &= \mathbf{C}_{{\gamma}_0} \left( \frac{1}{\phi}
        \sum_{t,\ell} (\tilde{\mathbf{Y}}_{t\ell} -  \mathbf{X}_+\, \bm{\gamma})
        + \frac{1}{\sigma_{{\gamma}}^2}\mathbf{R}^{-1}_{\phi_{{\gamma}}}  \tilde{\mathbf{X}} \tilde{\bm{\gamma}} \right),
    \end{align*}
    where $\bm\Omega = \text{diag}(n_{1}^{+}, \dots, n_n^{+})$, 
    with $n_i^{+}$ as the number of days with positive true precipitation values in station $\bs_i$.
    
    Given the normal prior, $\bm{\gamma} \sim N_{p-k}(\mathbf{a}_{{\gamma}}, \mathbf{B}_{{\gamma}})$, we sample the values of $\bm\gamma$ from their non-closed form full posterior 
    \begin{align*}
        [\bm{\gamma} \mid \mathbf{Y}^{\text{true}}_+, \bm\gamma_0, \phi] 
        &\propto 
        \exp\left\{ -\frac{1}{2} (\bm{\gamma} - \mathbf{a}_{{\gamma}})^\top \mathbf{B}_{{\gamma}}^{-1} (\bm{\gamma} - \mathbf{a}_{{\gamma}}) -
        \frac{1}{\phi} \mathbf{1}_{N_+}^\top \left( \tilde{\mathbf{Y}} - \bm{\gamma}_0 + \mathbf{1}_{N_+} \right) \right\}.
    \end{align*}
    
    We update the regression coefficients following the same procedure as with $\bm\gamma_0$, where the candidate $\bm{\gamma}^*$ is drawn from the following normal proposal distribution,
    \begin{align*}
        q(\bm{\gamma}^* \mid \bm{\gamma}) &= N_{p - k}(\bm{\gamma}^* \mid \mathbf{m}_{{\gamma}} , \mathbf{C}_{{\gamma}} ); \\
        \mathbf{C}_{{\gamma}} &= \left( \frac{1}{\phi} \mathbf{X}_+^\top \mathbf{X}_+ + \mathbf{B}_{{\gamma}}^{-1} \right)^{-1}, \\
        \mathbf{m}_{{\gamma}} &= \mathbf{C}_{{\gamma}} \left( \frac{1}{\phi} \mathbf{X}_+^\top 
        (\tilde{\mathbf{Y}} - \bm{\gamma}_0)
        + \mathbf{B}_{{\gamma}}^{-1} \mathbf{a}_{\gamma} \right).
    \end{align*}
    
    Given an improper flat prior for the dispersion parameter $\phi$ on the log scale, i.e., $[\log(\phi)] \propto 1$, we sample values for $\log(\phi)$ from its full conditional,
    {\small
    \begin{align*}
        [&\log(\phi) \mid \mathbf{Y}^{\text{true}}, \bm{\gamma}, \bm{\gamma}_0] \propto  
        \exp\left\{- \frac{1}{\phi} \Big[ \mathbf{1}^\top_{N_+} \left( \tilde{\mathbf{Y}} + \mathbf{1}_{N_+} - \log(\mathbf{Y}^{\text{true}}_+) \right) + N_{+} \big( \log(\phi) + \phi \log(\Gamma(1 / \phi)) \big) \Big] \right\}.
    \end{align*}
    }
    We update this parameter using a random-walk Metropolis step. This entails using a normal proposal with a variance parameter tuned to achieve an acceptance rate of approximately $33$\% \citep{roberts2009}.

    \paragraph{Full conditional distribution of the Gaussian process hyperparameters.} 
    For the spatially varying coefficients, we also need to sample the hyperparameters corresponding to the mean regression coefficients, the variance, and the decay parameter. The procedure for the hyperparameters of $\gamma_0(\bs)$ is shown below, but it is analogous for those of $\beta_0(\bs)$. Given the prior distribution of the hyperparameters: $\tilde{\bm{\gamma}} \sim N_k \left(\mathbf{a}_{\tilde{{\gamma}}}, \mathbf{B}_{\tilde{{\gamma}}} \right)$, $1/\sigma^2_{\gamma} \sim G(a_\sigma,b_\sigma)$, and   $\phi_{\gamma} \sim G(a_\phi,b_\phi)$; the full posterior distributions are:
    \begin{align*}
        &\tilde{\bm{\gamma}} \mid \dots \sim N_k (\mathbf{m}_{\tilde{{\gamma}}}, \mathbf{C}_{\tilde{{\gamma}}}), \\ 
        & \mathbf{C}_{\tilde{{\gamma}}} = \left(\frac{1}{\sigma_{\gamma}^2} \mathbf{1}_n^\top \mathbf{R}^{-1}_{\phi_{\gamma}} \mathbf{1}_n + \mathbf{B}_{\tilde{{\gamma}}}^{-1}\right)^{-1}, \\
        & \mathbf{m}_{\tilde{{\gamma}}} = \mathbf{C}_{\tilde{{\gamma}}} \left( \frac{1}{\sigma_{\gamma}^2} \mathbf{1}_n^\top \mathbf{R}^{-1}_{\phi_{\gamma}} \bm{\gamma}_0 + \mathbf{B}_{\tilde{{\gamma}}}^{-1} \mathbf{a}_{\tilde{{\gamma}}} \right);
    \end{align*}
    and    
    \begin{equation*}
    1/\sigma_{\gamma}^2 \mid \dots \sim G\left( a_\sigma + \frac{n}{2}, b_\sigma + \frac{1}{2}(\bm{\gamma}_0 -  \tilde{\mathbf{X}} \tilde{\bm{\gamma}})^\top \mathbf{R}^{-1}_{\phi_{\gamma}}(\bm{\gamma}_0 -  \tilde{\mathbf{X}} \tilde{\bm{\gamma}})
    \right).
    \end{equation*}
    The decay parameter does not have a standard full conditional distribution; therefore, an adaptive random-walk Metropolis step on the logarithmic scale \citep{roberts2009}, tuned to a 33\% acceptance rate, is employed with the following kernel:
    \begin{align*}
        [\log\phi_{\gamma} \mid \dots ]
        &\propto 
        \lvert \mathbf{R}_{\phi_{\gamma}} \rvert^{-1/2} \cdot 
        \exp\left\{ -\frac{1}{2\sigma_{\gamma}^2}(\bm{\gamma}_0 - \tilde{\mathbf{X}} \tilde{\bm{\gamma}})^\top \mathbf{R}_{\phi_{\gamma}}^{-1} (\bm{\gamma}_0 - \tilde{\mathbf{X}} \tilde{\bm{\gamma}}) + a_\phi\log\phi_{\gamma} - b_\phi \phi_{\gamma} \right\}.
    \end{align*}

\paragraph{Inference for $\bm{\epsilon}$.}
For simplicity, we first consider the case of a common $\epsilon$. If we wish to draw inference about it, the constraint imposed by the likelihood is simply:
\begin{equation*}
   a_{\epsilon} = \max_{\substack{1 \le t \le T \\ 1 \le \ell \le L \\ 1 \le i \le n}} \left\{ Y_{t\ell}^{\text{true}}(\bs_i) : y_{t\ell}^{\text{obs}}(\bs_i) = 0 \right\} < \epsilon \le \min_{\substack{1 \le t \le T \\ 1 \le \ell \le L \\ 1 \le i \le n}} \left\{ y_{t\ell}^{\text{obs}}(\bs_i) : y_{t\ell}^{\text{obs}}(\bs_i) > 0 \right\} = b_{\epsilon}.
\end{equation*}
Then, the full conditional distribution of $\epsilon$ is a truncated prior distribution:
\begin{equation*}
    [\epsilon \mid \mathbf{y}^{\text{obs}},\mathbf{Y}^{\text{true}},\mathbf{Z},\bm{\theta}_{-\epsilon}] \propto \mathbf{1}\{a_{\epsilon} < \epsilon \le b_{\epsilon}\} \cdot [\epsilon].
\end{equation*}
The prior distribution can be continuous or defined over a set of known discrete values.

In the case of varying thresholds $\epsilon_{t\ell}(\bs_i)$, the increased parameter dimensionality necessitates a more informative prior to ensure identifiability. A promising avenue is to allow these thresholds to take discrete values dictated by a change-point mechanism, which can be elegantly modeled via a Dirichlet process to automatically infer the number of regimes.

\bibliography{bibliography.bib}

\end{document}